\documentclass[11pt]{article}
\usepackage{jheppub}
\usepackage{subfigure}
\usepackage{rotating}

\newcommand{\beq}{\begin{eqnarray}}
\newcommand{\eeq}{\end{eqnarray}}
\newcommand{\non}{\nonumber\\}

\newcommand{\Tr}{{\rm Tr}}
\newcommand{\diag}{{\rm diag}}

\newcommand{\Li}{{\rm Li}}
\newcommand{\h}[2]{h\left(#1,#2\right)}
\newcommand{\iotafunc}[4]{\iota\left(#1,#2,#3,#4\right)}
\newcommand{\sigmafunc}[4]{\sigma\left(#1,#2,#3,#4\right)}

\title{Mediation of Supersymmetry Breaking in Quivers}

\author{Roberto Auzzi,}
\author{Amit Giveon}
\author{and Sven Bjarke Gudnason}

\affiliation{Racah Institute of Physics, The Hebrew University,
 Jerusalem 91904, Israel}

\emailAdd{auzzi(at)phys.huji.ac.il}
\emailAdd{giveon(at)phys.huji.ac.il}
\emailAdd{gudnason(at)phys.huji.ac.il}

\abstract{The soft masses due to SUSY breaking,
mediated by gauge fields, are computed
for generic matter in quiver gauge theories.}

\begin{document}
\maketitle

\section{Introduction}

The mediation of supersymmetry (SUSY) breaking in quiver gauge
theories has various interesting properties.
First, attaching the SUSY-breaking sector and the matter superfields
to different nodes of the quiver gives rise to a suppression of the
sfermion masses, thus allowing a sufficiently light stop to explain
the hierarchy problem, and produces exotic sparticle spectra
with interesting collider signatures
\cite{DeSimone:2008gm,DeSimone:2009ws}.
This is especially important in the dynamical embedding
of SUSY breaking and its mediation to the Minimal Supersymmetric
extension of the Standard Model (MSSM) \cite{Green:2010ww},
where the gaugino masses vanish to leading order in SUSY breaking
\cite{Komargodski:2009jf}.
Moreover, separating the matter fields on different
nodes gives rise to Yukawa-coupling textures which
may deal with the flavor puzzle of the SM \cite{Craig:2011yk}.
Quiver gauge theories also appear in four-dimensional
low-energy realizations of models with large extra dimensions
\cite{ArkaniHamed:2001ca}
-- viz.~in gaugino mediated SUSY-breaking models
\cite{Kaplan:1999ac,Chacko:1999mi,Csaki:2001em,Cheng:2001an} --
thus providing further motivation to investigate
the mediation of SUSY breaking in quivers.

Recent studies of SUSY breaking and its mediation in the minimal case
-- a quiver with two nodes -- already led to
rather surprising results.
First, one finds that both the right-handed
and the left-handed sleptons can be lighter than the bino in the
low-scale mediation regime \cite{DeSimone:2008gm,DeSimone:2009ws},
even when the messenger scale is comparable
to the masses of the additional gauge particles
\cite{Auzzi:2010mb,Auzzi:2010xc}.
Moreover, in this hybrid low-scale mediation case,
the sfermion masses are comparable
to those of the gauginos even when the latter
vanish to leading order in SUSY breaking \cite{Auzzi:2011gh}.
Finally, already this minimal setting provides
intriguing ways to tackle the flavor puzzle
as well as the $\mu/B\mu$ problem \cite{Craig:2011yk}.

In this work, we compute the soft SUSY-breaking masses in a large class of models.
Explicitly, in section \ref{sec:setting}, we set up our supersymmetric
quivers, with matter and messengers in generic representations of the
quiver's product gauge groups, and present the general form of the soft
masses.
When all the MSSM matter fields are charged under the same node of the
quiver, the theory is flavor blind. However,
here we find the soft masses also for the generic case --
when the matter fields are distributed on {\it different} nodes --
in which case the flavor texture is rich.

The details of each model are encoded in a unique form factor,
as we prove in section \ref{sec:formfactors}, where we also compute
its value.
In section \ref{sec:examples}, we analyze in more detail several
examples of particular interest. Our emphasis is on models that may
have perturbative unification, and we thus focus on quivers
with at most five nodes.
We inspect in more detail various examples of quivers with three
nodes, which may incorporate the main freedom in flavor textures.

In the hybrid case -- when the various scales in the problem are
comparable -- we find the following main property.
The suppression of the scalar masses is significant
when the matter superfields and messengers are not charged
under the same node of the quiver,
for any value of the messenger scale.
This opens up phenomenologically appealing avenues,
which we discuss in section \ref{sec:discussion}.
Finally, in the appendix we present some technical details.

\section{Setting}\label{sec:setting}

\begin{figure}[!htp]
\begin{center}
\includegraphics[width=0.6\linewidth]{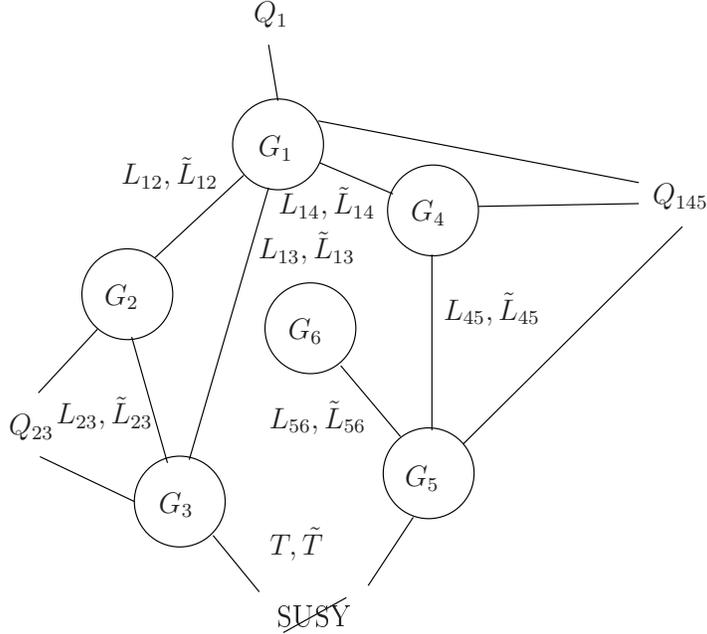}
\caption{A random quiver representing a theory for which we
  compute the soft masses.}
\label{fig:randomquiver}
\end{center}
\end{figure}

In this note we consider mediation of SUSY breaking in a
generic class of quiver gauge theories with $N$ nodes connected by
$K\geq N-1$ bifundamental link fields and coupled to messenger fields
in arbitrary representations, which are charged under at least one of
the nodes; see fig.~\ref{fig:randomquiver}.
Each node represents a gauge group $G_i$, $i=1,\ldots,N$, which we
take, for simplicity, to be all the same group.\footnote{For
  phenomenological purposes one would be interested in the case
  where at low energies one of the groups is the SM one, say
  $G_1=SU(3)\times SU(2)\times U(1)$. If one also imposes unification
  then the rest of the nodes need to be $G_{\rm GUT}$-invariant,
  e.g.~$G_i=SU(5)$ for $i>1$. Perturbative unification further imposes an
  upper bound on the number of nodes.  Our setting and
  formalism is generalized straightforwardly to such a case by summing
  over each of the SM subgroup factors in each node with their
  corresponding gauge couplings. }
Each line linking two gauge groups represents
a pair of bifundamental and anti-bifundamental
chiral superfields $(L_{ij}, \tilde{L}_{ij})$,
whose vacuum expectation values (VEVs) break the gauge symmetry
to a linear combination of the various $G_i$'s.
The superpotential of the link fields sector is given by
\beq
\mathcal{W}_L = \sum_{I=\{ij\}}^{K} H_{ij} \,
\left(\Tr(L_{ij}\tilde{L}_{ij}) - v_{ij}^2\right) + \mathcal{W}_A \ ,
\eeq
where $H_{ij}$ are singlet superfields and
$v_{ij}$ are the VEVs of the link fields
($\langle L_{ij}\rangle = \langle\tilde{L}_{ij}\rangle =
v_{ij}\mathbf{1}$).
Ideally, the VEVs are dynamically generated by some theory at higher
energies, however, for simplicity, we shall impose them by hand (this
does not change our analyzes in the following).
The superpotential $\mathcal{W}_A$ includes terms that give mass to
the combinations of link fields which are not eaten by the gauge
fields and which are not in the same $\mathcal{N}=1$ multiplet
as the massive vector boson.
This can be achieved for instance with an adjoint field $A_i$ of one
of the groups $G_i$ under which the set of link fields is charged
\cite{Cheng:2001an},
\beq \mathcal{W}_A=
 \sum_{I=\{ij\}}^K \Tr(L_{ij} A_{i} \tilde{L}_{ij}) \ .
\eeq
The details of this mechanism are not important for the calculation of
the soft masses.

We denote by $A^i_\mu$ the gauge field of the group $G_i$, and by
$g_i$ the corresponding gauge coupling.
The unbroken combination $\tilde{G}$ of the $G_i$'s, which in
phenomenological applications is identified with the Standard Model
(SM) gauge group, is the following,
\beq
\tilde{A}_\mu =
\frac{\sum_i^N \prod_{j\neq i}^N g_j A_\mu^i}{\sqrt{P_{N-1}(\{g_k^2\})}}
\ ,\label{aaa}
\eeq
where $P_{N-1}(\{x_i\})\equiv \sum_i^N \prod_{j\neq i}^N x_j$ is the
$(N-1)$-th symmetric polynomial.
Note that the linear combination (\ref{aaa}) is independent of the VEVs.
The gauge coupling of this unbroken combination is
 \beq
 \frac{1}{g_{\rm eff}^2} = \sum_{i=1}^{N} \frac{1}{g_i^2} \ .
 \eeq
The scale of the VEVs is taken to be sufficiently larger than the
electroweak scale, such that all the matter which gets a mass from the
Higgsing of the link fields can be considered decoupled.

We denote the matter fields by $Q_{i_1i_2\cdots i_{P_A}}^A$,
which are chiral superfields transforming under the representation
$(\mathbf{r}_{i_1}^A,\mathbf{r}_{i_2}^A,\ldots,\mathbf{r}_{i_{P_A}}^A)$
of the gauge group
$G_{i_1}\times G_{i_2}\times\cdots\times G_{i_{P_A}}$, where $A$
labels the matter fields and $P_A\leq N$ is the number of groups under
which the field $Q^A$ is charged.
Each MSSM matter field is charged under one
of the $G_i$'s; however, our formalism applies also to soft masses in
more general settings with various exotic matter fields
as well as to soft masses of the link fields.

\begin{figure}[!htp]
\begin{center}
\mbox{\subfigure[]{\includegraphics[width=0.22\linewidth]{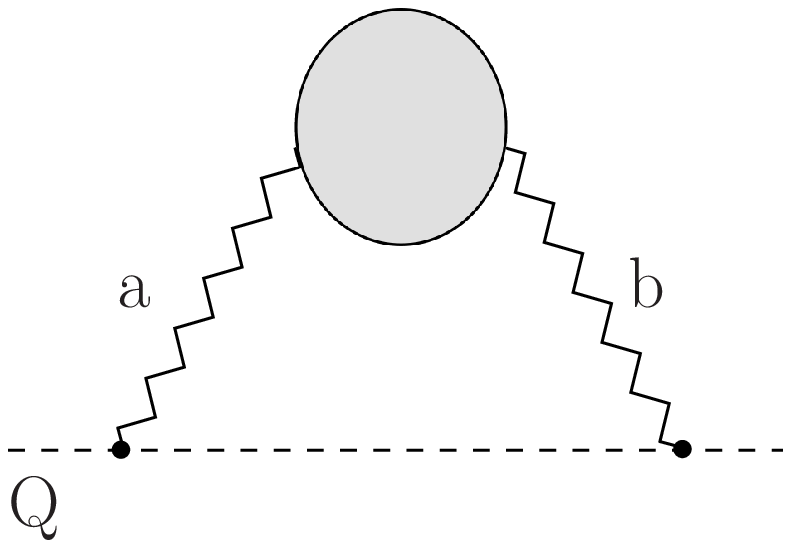}}\quad
\subfigure[]{\includegraphics[width=0.22\linewidth]{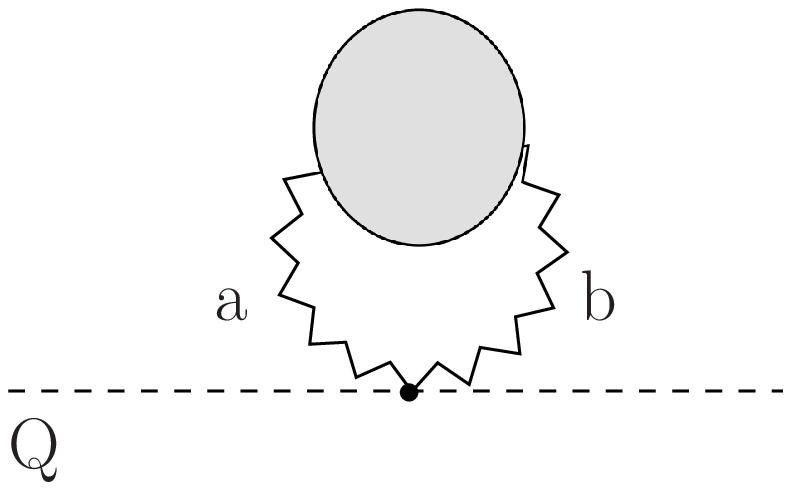}}\quad
\subfigure[]{\includegraphics[width=0.22\linewidth]{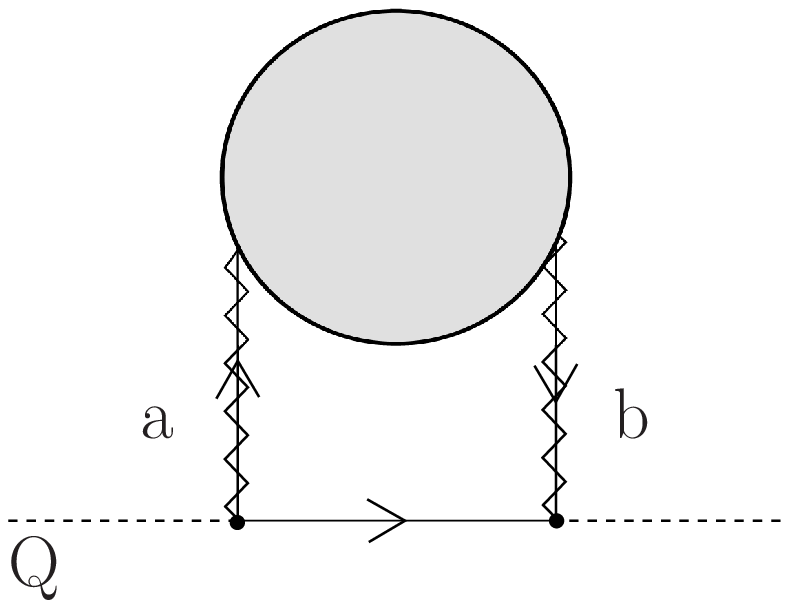}}\quad
\subfigure[]{\includegraphics[width=0.22\linewidth]{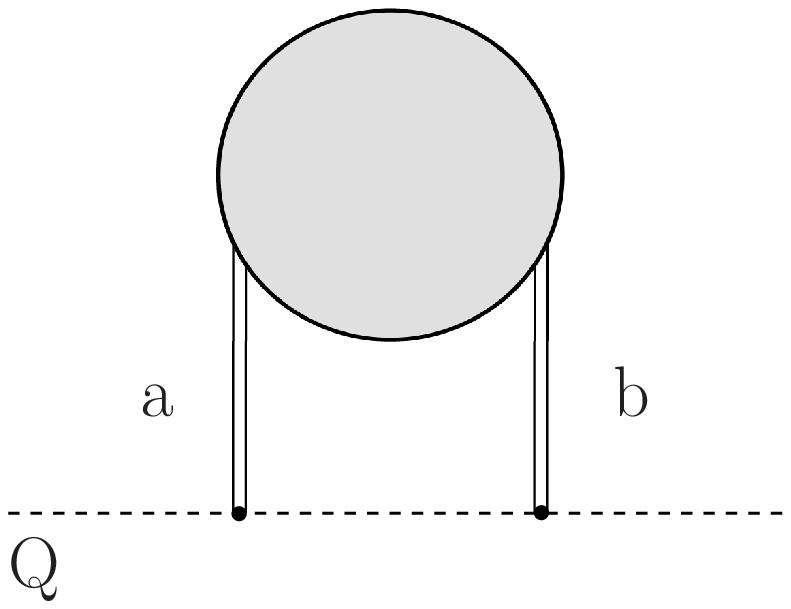}}}
\caption{Contributions to the sfermion masses due to (a,b) gauge
  bosons, (c) gauginos and (d) D-terms.}
\label{fig:diagrams}
\end{center}
\end{figure}

We assume that SUSY is broken in some hidden sector and that the SUSY
breaking is communicated to the visible sector by the gauge
interactions of a subset of the groups $G_i$.
In order to perform the following analysis in its full generality, it
is convenient to use the global current multiplet formalism of
\cite{Meade:2008wd,Buican:2008ws}.
Let us consider a current $j_\mu^{t,m}$ charged under $G_{t}$, which
is embedded in a real superfield $\mathcal{J}^{t,m}$ containing also a
scalar $J^{t,m}$ component and a spinor $j_\alpha^{t,m}$, where
$m=1,\ldots,\dim(G_{t})$ is the adjoint index of $G_{t}$. The
functions $C_r^t(x),B_{1/2}^t(x)$ parametrize the current correlators
as follows
\begin{eqnarray}
\left\langle J^{t,m}(x)J^{t,n}(0)\right\rangle &\equiv&
  C_{0}^t(x) \delta^{mn} \ , \non
\big\langle j_{\alpha}^{t,m}(x) \bar{j}_{\dot{\beta}}^{t,n}(0)\big\rangle
  &\equiv&
  -i\sigma_{\alpha\dot{\beta}}^{\mu}\partial_{\mu}C_{1/2}^t(x)
  \delta^{mn} \ , \label{correnti} \\
\left\langle j_{\mu}^{t,m}(x)j_{\nu}^{t,n}(0)\right\rangle &\equiv&
 \left(\eta_{\mu\nu}\partial^2 - \partial_{\mu}\partial_{\nu}\right)
  C_{1}^t(x) \delta^{mn} \ , \nonumber \\
\big\langle j_{\alpha}^{t,m}(x)j_{\beta}^{t,n}(0)\big\rangle &\equiv&
 \frac{1}{4} \epsilon_{\alpha \beta} B_{1/2}^t(x) \delta^{mn} \ ,
 \nonumber
\end{eqnarray}
where $\alpha,\beta,\dot{\beta}=1,2$ are spinor indices and the index
$t$ is referring to the group $G_t$. Now we can couple the
SUSY-breaking current to $R$ of the gauge groups
$G_{t_1}\times G_{t_2}\times\cdots\times G_{t_R}$.
We denote by $\tilde{C}_r^t(p), \tilde{B}_{1/2}^t(p)$ the Fourier
transforms of $C_r^t(x), B_{1/2}^t(x)$. The functions
$\tilde{C}_1^t(p),\tilde{C}_{1/2}^t(p),\tilde{C}_0^t(p)$
parametrize the contribution to the sfermion masses mediated by
gauge bosons (fig.~\ref{fig:diagrams}a,b), gauginos
(fig.~\ref{fig:diagrams}c) and scalars (fig.~\ref{fig:diagrams}d),
respectively, viz.~they represent the various blobs of the
figure.
Contributions to (the light) gaugino masses are instead parametrized
by $\tilde{B}_{1/2}^t(p)$.

If the theory contains a weakly coupled messenger sector, the
components of the supercurrent $\mathcal{J}^{t_u,m}$, $u=1,\ldots,R$,
are constructed in terms of the SUSY-breaking messengers fields
$T_B,\tilde{T}_{B}^\dag$,
$B=1,\ldots,n_{\rm mess}$, which transform under the representations
$(\mathbf{s}_{t_1}$, $\mathbf{s}_{t_2},\ldots,\mathbf{s}_{t_{R}})$ of
the gauge group $G_{t_1}\times G_{t_2}\times\cdots\times G_{t_{R}}$,
and $R\geq 1$ is the number of groups under which the messengers are
charged.
In this case, the components of $\mathcal{J}^{t,m}$ can be written
explicitly
\begin{eqnarray}
J^{t,m} &=& T_{B}^{*} t_{t}^m T_{B}
 - \tilde{ T_{B}}^{*} t_{t}^m \tilde{T}_{B} \ , \non
j_{\alpha}^{t,m} &=& -\sqrt{2}i\left(T_{B}^{*} t_{t}^m \psi_{T_{B}\alpha}
  - \tilde{T_{B}}^* t_{t}^m \psi_{\tilde{T_B}\alpha}\right)  \ , \\
j_{\mu}^{t,m} &=& i\left(T_{B} t_{t}^m \partial_{\mu} T_{B}^{*}
  -T_{B}^{*} t_{t}^m \partial_{\mu}T_{B}
  -\tilde{T}_{B} t_{t}^m \partial_{\mu}\tilde{T}_{B}^{*}
  +\tilde{T}_{B}^{*} t_{t}^m \partial_{\mu}\tilde{T}_{B} \right) \non &&\;
  +\,\psi_{T_B}\sigma_{\mu} t_{t}^m \bar{\psi}_{T_B}
  -\psi_{\tilde{T}_B}\sigma_{\mu} t_{t}^m \bar{\psi}_{\tilde{T}_B} \ ,
  \nonumber
\end{eqnarray}
where $t_{t}^m$ are the generators of $G_t$.

In the explicit examples in section \ref{sec:examples}, for
simplicity, we will focus on a minimal sector with just a single
messenger coupled to an F-term spurion of SUSY breaking, $S$, via the
superpotential $\mathcal{W}_T= \lambda_S S T \tilde{T}$.
Explicit expressions for the functions
$\tilde{C}_r^t(p),\tilde{B}_{1/2}^t(p)$ in the case of a general
weakly coupled messenger sector, coupled both to an F-term and a D-term
spurion, can be found in \cite{Marques:2009yu} and in
Appendix B.5 of \cite{Dumitrescu:2010ha}.

The gauginos of the unbroken gauge group $\tilde{G}$ acquire a
SUSY-breaking soft mass at one loop, which can be computed as in gauge
mediation \cite{Meade:2008wd},
\beq
M_{\tilde{g}}=\frac{g_{\rm eff}^2}{4} \, \sum_{u=1}^R
\tilde{B}_{1/2}^{t_u}(p^2 = 0 ) \ .
\eeq
If additional matter fields in the adjoint representation are present,
the gauginos can also acquire Dirac masses (see
e.g.~\cite{Benakli:2008pg,Abel:2011dc} for a recent discussion).

The purpose of this note is to find the value for the scalar soft
masses of the matter field $Q_{i_1 i_2\cdots i_{P_A}}^A$; the result is
\begin{align}
m_{i_1 i_2\cdots i_{P_A}}^2 = - g_{{\rm eff}}^4
\sum_{j=1}^{P_A} \sum_{u=1}^{R} & \;
c_{2} \big(\mathbf{r}_{i_j}^A\big)
\int \frac{d^4p}{(2\pi)^4}\frac{1}{p^2}
f_{i_j}^{t_u}(p^2)
\left[
3\tilde{C}_1^{t_u}(p) - 4\tilde{C}_{1/2}^{t_u}(p) + \tilde{C}_0^{t_u}(p)
\right]
\ , \label{eq:sfermion_mass}
\end{align}
where $c_{2} \big(\mathbf{r}_{i_j}^A\big)$ is the quadratic
Casimir of the representation under which the field
$Q_{i_1i_2\cdots i_{P_A}}^A$
transforms with respect to the group $G_{i_j}$ and the sum is over
all the groups under which the field is charged.
The  form factor $f_{i_j}^{t_u}(p^2)$  depends
on the quiver structure and the various parameters such as the gauge
couplings and the VEVs of the link fields, etc. $G_{i_j}$ is one of the
groups under which the matter is charged and $G_{t_u}$ is one of the
groups under which the messengers are charged.
For $N=1$, i.e.~the single node case, General Gauge Mediation (GGM) is
recovered \cite{Meade:2008wd}\footnote{In this case, in the limit of
  vanishing effective gauge coupling, $g_{\rm eff}\to 0$, the
  SM matter decouples from the SUSY-breaking sector, and thus the theory
  belongs to the GGM class, as defined in \cite{Meade:2008wd}.
  On the other hand, in the case of several nodes, the quiver theory does not
  necessarily obey this criterion. } and $f_{\rm GGM}=f_1^1=1$.
In the next section, we derive eq.~(\ref{eq:sfermion_mass})
and find the explicit expression for the form factors
$f_{i_j}^{t_u}(p^2)$.

Finally, the soft masses for the link fields, which are bifundamental fields,
are obtained from the same formula, \eqref{eq:sfermion_mass},
by summing over just two groups
(corresponding to $P_A=2$ and $i_1=i,i_2=j$ for a link field $L_{ij}$).

\section{The form factors}\label{sec:formfactors}

\subsection{Masses of the quiver degrees of freedom}

As a prelude, we discuss the masses in the quiver theory.
We parametrize the $K$ link fields using an index
$I=1,\ldots,K$; the link fields are denoted by $L_I (\tilde{L}_I$),
where each $I$ corresponds to a particular set of values $(i,j)$,
i.e., the fields transform as $(\square,\overline{\square})$
($(\overline{\square},\square)$) under the group $G_i\times G_j$.
For simplicity, we assume that all the link field VEVs $v_I$ are
real (in many cases this can be achieved by a gauge transformation;
cases where this cannot be achieved for all the VEVs would lead to
potentially dangerous CP violations).
We define the following $N\times K$ matrix
\beq
Z_{\ell I} \equiv \sqrt{2} g_\ell v_I
  \left(\delta_{\ell i} - \delta_{\ell j}\right) \ ,
\eeq
in terms of which the mass-squared matrix of the vector gauge bosons
is given by
\beq
\mathcal{M}_V^2 = Z Z^{\rm T} \ ,
\eeq
which is an $N\times N$ mass-squared matrix with $N-1$ non-vanishing
eigenvalues and one zero eigenvalue (the SM gauge
bosons). Diagonalization of the matrix is done by a unitary
transformation,
\beq
U \mathcal{M}_V^2 U^\dag = \mathcal{D}_V^2 =
\diag\left(m_1^2,m_2^2,m_3^2,\cdots,m_N^2\right) \ ,
\label{eq:Umatrix}
\eeq
where $D_V^2$ is a diagonal matrix with eigenvalues
$m_1^2\leq m_2^2\leq m_3^2\leq\cdots\leq m_{N-1}^2$, and $m_N=0$.

Only particular combinations of the link fields will participate
actively in the mediation of SUSY breaking; these degrees
of freedom are in the same $\mathcal{N}=1$ multiplets as the massive
gauge bosons and are given by
\beq
l_I \equiv \frac{1}{\sqrt{2}}{\rm Re}
  \left(\delta L_I - \delta\tilde{L}_I\right) \ ,
\eeq
where $L_I = v_I + \delta L_I$, and analogously for
$\tilde{L}_I$.
The mass-squared matrix of these fields is
\beq
\mathcal{M}_S^2 = Z^{\rm T} Z \ .
\eeq
Suppose now that $\mathrm{v}$ is an eigenvector of the matrix
$Z Z^{\rm T}$ with eigenvalue $\lambda$,
$Z Z^{\rm T} \mathrm{v} = \lambda \mathrm{v}$, then
$Z^{\rm T} \mathrm{v}$ is an eigenvector of $Z^{\rm T} Z$ with the
same eigenvalue $\lambda$.
Hence, for all eigenvectors of the mass-squared matrix for
the vector multiplet $\mathcal{M}_V^2$ with non-vanishing eigenvalue
$\lambda\neq 0$, there exist the same
mass-squared for the scalars (as SUSY dictates). Due to the fact
that there are at most $N-1$ non-vanishing eigenvalues of both
mass-squared matrices, we can diagonalize the $K \times K$ mass-squared
matrix $\mathcal{M}_S^2$ as follows
\beq
W \mathcal{M}_S^2 W^\dag = \mathcal{D}_V^2  \ ,
\eeq
where $W$ is an $N\times K$ matrix.

The gauginos $\lambda_i$ are mixed with some linear combinations of the
link fields by the following Dirac mass term,
\beq
Z_{\ell I} \lambda_\ell \psi_{l_I} + {\rm h.c.}\ ,
\eeq
where
$\psi_{l_I}=\frac{1}{\sqrt{2}}(\psi_{L_I}-\psi_{\tilde{L}_I})$.
These mass terms can be diagonalized as follows,
\beq
U Z W^\dagger = \mathcal{D}_V \ ,
\eeq
which is formally given by $(\mathcal{D}_V^2)^{1/2}$.

\subsection{Computation of the form factors}

It is sufficient to compute the mass of a sfermion $Q_i$ charged under
a single group $G_i$, in presence of a SUSY-breaking sector charged
just under the group $G_t$.
The more general expression in eq.~(\ref{eq:sfermion_mass}) follows by
linearity.
There are three classes of diagrams to calculate, i.e.~the diagrams
involving gauge bosons, gauginos and scalars; see figure
\ref{fig:diagrams}.
Now we will show that all three classes of diagrams give rise to the
same form factor $f_{i}^{t}(p^2)$.

First, let us consider the diagrams involving gauge bosons; see
fig.~\ref{fig:diagrams}a,b.
The coupling of the vector bosons to the SUSY-breaking sector is
\beq
- g_t j_\mu^{t,m} A_t^{\mu m} \  ,
\eeq
where $j_\mu^{t,m}$ is the vectorial component of the current
multiplet presented near eq.~(\ref{correnti}).
The matter field $Q_i$, on the other hand, couples just to the gauge
boson $A_\mu^i$.
Both the gauge bosons $(A_\mu^t,A_\mu^i)$ are not mass eigenstates, in general;
the propagators of these fields should be decomposed into a
linear combination of the $N$ mass eigenstates.
Then the factor $g_{\rm eff}^2/p^2$  in the gauge mediation integrand
should be replaced by
\beq
g_i g_t \sum_{a=1}^N U_{ia}^\dag \frac{1}{p^2-m_a^2} U_{at} \ ,
\label{eq:prop_repl}
\eeq
where the matrix $U$ is the unitary matrix in eq.~\eqref{eq:Umatrix}
rotating to the mass eigenstates of the gauge bosons in which the
propagators are written.
The factor of eq.~\eqref{eq:prop_repl} is
independent of the blob of fig.~\ref{fig:diagrams}a,b and
hence the form factor reads
\beq
f_i^t(p^2) =
\left(\frac{g_i g_t}{g_{\rm eff}^2}\sum_{a=1}^N U_{ia}^\dag \frac{p^2}{p^2-m_a^2} U_{at}\right)^2
\ , \label{eq:formfactor}
\eeq
where both gauge boson $a$ and $b$ in the figure contribute with the
same factor and hence the result is a perfect square.

Now we turn to the diagram with the gauginos; see
fig.~\ref{fig:diagrams}c.
The coupling of the gauginos to the sfermions reads
\beq
g_i \left(Q_i t_i^m \bar{\psi}_{Q_i} \bar{\lambda}_i^m
-Q_i^* t_i^m \psi_{Q_i} \lambda_i^m \right) \ ,
\label{fermcoupling1}
\eeq
where $t_i^m$ are generators of the group $G_i$.
The coupling of gauginos to the SUSY-breaking sector, on the other hand,
is given by
\beq
- g_t \left(j^{t,m} \lambda_t^m + \bar{j}^{t,m} \bar{\lambda}_t^m
\right) \ ,
\eeq
where the spinor indices have been contracted and $j_\alpha^{t,m}$ is
the spinorial component of the current multiplet presented near
eq.~(\ref{correnti}). Again we should decompose the
propagators into mass eigenstates; the gauge mediation factor,
$i g_{\rm eff}^2\, p\cdot\sigma_{\alpha\dot{\beta}}/p^2$
in the integrand, is then replaced by
\beq
i g_i g_t
\sum_{a=1}^N U_{i a}^\dag
\frac{p\cdot\sigma_{\alpha\dot{\beta}}}{p^2-m_a^2}
U_{a t} \ ,
\eeq
which again gives rise to the same form factor
\eqref{eq:formfactor}.

Finally, we need to consider scalar-mediated diagrams; see
fig.~\ref{fig:diagrams}d.
The D-terms give rise to the trilinear couplings between the link
field scalars $l_I$ and the sfermions
\beq
g_i Z_{iI} l_I^m \left(Q_i^* t_i^m Q_i\right) \ , \label{3zampeA}
\eeq
while the SUSY-breaking sector similarly has the trilinear couplings,
\beq
g_t Z_{tI} l_I^m J^{t,m} \ , \label{3zampeB}
\eeq
where $J^{t,m}$ is the scalar component of the current multiplet
presented near eq.~(\ref{correnti}).
The D-terms provide also a direct coupling between the sfermions and
the SUSY-breaking sector,
\beq
g_i^2 \delta_{it} \left(Q_i^* t_i^m Q_i\right) J^{t,m} \ . \label{4zampe}
\eeq
The contribution due to the exchange of a scalar $l_I$
(which is due to the vertices in eqs.~\eqref{3zampeA},
\eqref{3zampeB})
should then again be decomposed in terms of mass eigenstate
propagators,
\begin{align}
g_i g_t \sum_{a=1}^N \sum_{I,J=1}^K
Z_{iI} W_{Ia}^\dag \frac{1}{p^2-m_a^2} W_{aJ} Z_{Jt}^{\rm T}
&= g_i g_t
U_{ij}^\dag U_{jk} Z_{kI} W_{Ia}^\dag \frac{1}{p^2-m_a^2}
W_{aJ} Z_{Jl}^{\rm T} U_{lm}^\dag U_{mt} \non
&= g_i g_t \sum_{a=1}^N
U_{ia}^\dag \frac{m_a^2}{p^2-m_a^2} U_{at} \ ,
\end{align}
where we have used that
$U Z W^\dag = \mathcal{D}_V = (\mathcal{D}_V^2)^{1/2}$ is the diagonal
$N\times N$ mass matrix. In the case that $i=t$ we still have the
contribution from the vertex in eq.~(\ref{4zampe}), hence, writing
down the total form factor we obtain
\beq
f_i^t(p^2) =
\left(\frac{g_i g_t}{g_{\rm eff}^2} \sum_{a=1}^N
\left(U_{ia}^\dag \frac{m_a^2}{p^2-m_a^2} U_{at} + \delta_{it}\right)
\right)^2
\ ,
\eeq
which is exactly that of eq.~\eqref{eq:formfactor}.
This establishes the proof of eq.~\eqref{eq:sfermion_mass} and
provides the explicit value of the form factor $f_i^t(p^2)$.

\section{Examples} \label{sec:examples}

In this section we shall focus on a minimal messenger sector,
with just a single messenger pair
coupled to an F-term spurion $S$ via the superpotential
\beq
\mathcal{W}_T= S T \tilde{T} \ , \qquad
\langle S \rangle = M + \theta^2 F \ . \label{wmgm}
\eeq
In this case the current correlators are given by
eqs.~\eqref{eq:C0}-\eqref{eq:C1} in the appendix.

Once $f_{i_j}^{t_u}(p^2)$ are given, the integrals can be directly
evaluated given the SUSY-breaking contributions parametrized by the
current correlators $\tilde{C}_r^t(p)$. It will be convenient to write
the scalar masses for $Q^A$ in the following form
\beq
m_{i_1i_2\cdots i_{P_A}}^2 =  2 \left(\frac{\alpha_{\rm eff}}{4\pi}\right)^2
\left(\frac{F}{M}\right)^2  \,
\sum_{j=1}^{P_A}
\sum_{u=1}^{R}
c_{2} \big(\mathbf{r}_{i_j}^A\big) \,
n\big(\mathbf{s}_{t_u}\big)
\mathcal{E}_{i_j}^{t_u}(x,\{y_\ell\}) \ ,
\label{eq:sparticle_masses}
\eeq
where $\alpha_{\rm eff}\equiv\frac{g_{\rm eff}^2}{4\pi}$,
$M$ is the messenger scale and $F/M$
is the effective SUSY-breaking scale.
$c_{2} \big(\mathbf{r}_{i_j}^A\big)$ is the quadratic
Casimir of the representation under which the field
$Q_{i_1i_2\cdots i_{P_A}}^A$ transforms with respect to the group
$G_{i_j}$ and the sum is over all the groups under which
this field is charged.
$n\big(\mathbf{s}_{t_u}\big)$ is the Dynkin index of the
representation under which the messenger fields $T_B,\tilde{T}_B^\dag$
transform with respect to the group $G_{t_u}$
(e.g.~$n = 1$ for a single set of ${\bf r}+\bar{\bf r}$ of $SU(r)$),
and the sum is over the groups under which the messenger fields
$T_B,\tilde{T}_B^\dag$ are charged.
Finally, we have defined a general function
$\mathcal{E}_{i_j}^{t_u}(x,\{y_\ell\})$ for matter charged under the gauge group
$G_{i_j}$ and a messenger charged under the gauge group $G_{t_u}$.
This function provides the measure of suppression of the sfermion masses
relative to the Minimal Gauge Mediation (MGM) case,
and we shall consequently refer to it as
``the sfermion mass suppression function.''
The variables in the suppression function are defined as
\beq
x \equiv \frac{F}{M^2} \ , \qquad
y_\ell \equiv \frac{m_{\ell}}{M} \ , \qquad \ell=1,\ldots,N-1 \ ,
\label{eq:variables}
\eeq
where $m_\ell$ is the $\ell$th mass eigenvalue of
the gauge bosons in the diagonal basis of eq.~\eqref{eq:Umatrix}.

In the following subsections we consider various examples.
As our first example, we will review the two nodes quiver
\cite{McGarrie,Auzzi:2010mb,Sudano:2010vt}.
For three nodes quivers,
we will consider all possible models,
which may be useful in applications to the SM flavor texture.
Finally, for four and five nodes, we will
present some properties of the linear quiver, which appears e.g.
in deconstructing extra dimensional models of gaugino mediation
\cite{Csaki:2001em,Cheng:2001an}.

\subsection{Two nodes quiver}\label{sec:twonodes}

Here we consider the two nodes quiver, which is shown in
fig.~\ref{fig:linear_quiver} for $N=2$.
For this quiver the mass-squared matrix reads
\beq
\mathcal{M}_V^2 = 2 v_{12}^2
\begin{pmatrix}
g_1^2 & -g_1 g_2  \\
-g_1 g_2 & g_2^2
\end{pmatrix} \ ,
\eeq
which includes the contribution from both
$L_{12},\tilde{L}_{12}$.\footnote{We normalize the trace of the
  generators of the group $G_i$ in the standard way
  $\Tr(t_i^m t_i^n) = \delta^{m n}/2$. }
The eigenvalues of the matrix are $0$ and
$m_1^2 = 2(g_1^2+g_2^2)v_{12}^2$, corresponding to
the mass-squared of the MSSM gauge bosons and the massive ones,
respectively.
Let us first review the results of \cite{Auzzi:2010mb}, where the
soft masses were calculated for chiral superfields
charged under the first node, $G_1$.
To use the general formula \eqref{eq:sfermion_mass},
we first need the diagonalization matrix $U$ of
eq.~\eqref{eq:Umatrix}, which in this case is given by
\beq
U = \frac{1}{\sqrt{g_1^2+g_2^2}}
\begin{pmatrix}
g_1 & -g_2 \\
g_2 & g_1
\end{pmatrix} \ .
\eeq
Plugging this matrix into eq.~\eqref{eq:formfactor} and using the
eigenvalues of the mass-squared matrix reproduces the form factor
\cite{McGarrie,Auzzi:2010mb,Sudano:2010vt}
\beq
f_1^2(p^2) = \frac{m_1^4}{\left(p^2-m_1^2\right)^2} \ , \qquad
g_{\rm eff}^2 = \frac{g_1^2g_2^2}{g_1^2 + g_2^2} \ .
\label{eq:2nodes_formfactor1}
\eeq
Evaluating the integral of eq.~\eqref{eq:sfermion_mass} using the
current correlators \eqref{eq:C0}-\eqref{eq:C1} in appendix
\ref{app:integrals}, gives the following result
\begin{align}
\mathcal{E}_1^2(x,y) &= \frac{1}{x^2} \left[
\alpha_0(x) - \alpha_1(x,y) - y^2 \alpha_2(x,y)
-\frac{2}{y^2}\beta_{-1}(x) + \beta_0(x) + \frac{2}{y^2} \beta_1(x,y)
+\beta_2(x,y) \right] \ ,
\end{align}
where $y\equiv y_1$, the suppression function is defined in
eq.~\eqref{eq:sparticle_masses} and the functions
$\alpha,\beta$ are defined in
eqs.~\eqref{eq:alpha0}-\eqref{eq:beta2}.
It is clear from eq.~\eqref{eq:2nodes_formfactor1} that the sfermion
mass becomes that of MGM
\eqref{eq:Efunc_MGM} for $y\to\infty$ and goes to zero in the limit of
$y\to 0$ as $\mathcal{E}_1^2(x,y) \approx y^2/6 + \mathcal{O}(y^3)$
for small $x$ (see appendix \ref{app:limits} for details about the
limits of the functions $\alpha,\beta$).

For matter charged under the second node, $G_2$, a similar calculation using
eq.~\eqref{eq:formfactor} gives
\begin{align}
f_2^2(p^2) &=
\left(\frac{\lambda_2 p^2-m_1^2}{p^2-m_1^2}\right)^2 \ , \qquad
\lambda_2 \equiv \frac{g_2^2}{g_{\rm eff}^2} \ .
\label{eq:2nodes_formfactor2}
\end{align}
Again, evaluating the integral of eq.~\eqref{eq:sfermion_mass} yields
\begin{align}
\mathcal{E}_2^2(x,y,\lambda_2) &= \frac{1}{x^2}\bigg[
\alpha_0(x)
-\left(1-\lambda_2^2\right)\alpha_1(x,y)
-(1-\lambda_2)^2 y^2 \alpha_2(x,y)
-\frac{2(1-\lambda_2)}{y^2}\beta_{-1}(x)
+\beta_0(x)
\non & \phantom{=\frac{1}{x^2}\bigg[\ }
+\frac{2(1-\lambda_2)}{y^2}\beta_1(x,y)
+(1-\lambda_2)^2\beta_2(x,y) \bigg] \ .
\end{align}
For this suppression function there are two limits which can be
understood
quite easily, viz.~the limit $y\to\infty$ yields the MGM result
\eqref{eq:Efunc_MGM}, as the Higgsing produces the
diagonal gauge group with the gauge coupling $g_{\rm eff}$,
while the opposite limit $y\to 0$ gives again the MGM result
\eqref{eq:Efunc_MGM}, but with gauge coupling $g_2$, instead.
Hence, in the limit $y\to 0$,
\beq
\mathcal{E}_2^2(x,0,\lambda_2) = \frac{\lambda_2^2}{x^2}\left[\alpha_0(x) +
  \beta_0(x)\right] \ .
\eeq
Note that in this limit the suppression function $\mathcal{E}_2^2$ is
larger than in the $y\to\infty$ limit,
since $\lambda_2^2 = (1+g_2^2/g_1^2)^2$
is greater than one for non-zero $g_2$.

\begin{figure}[btp]
\begin{center}
\mbox{\subfigure[]{\includegraphics[width=0.47\linewidth]{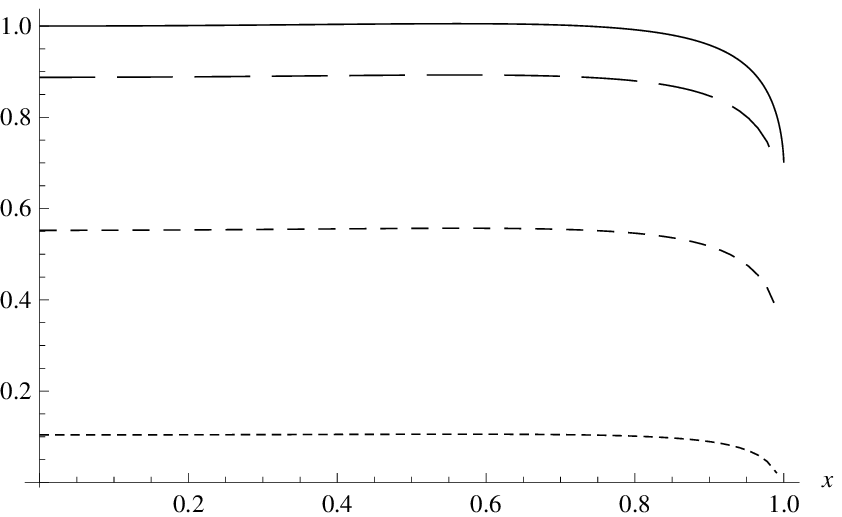}}\quad
\subfigure[]{\includegraphics[width=0.47\linewidth]{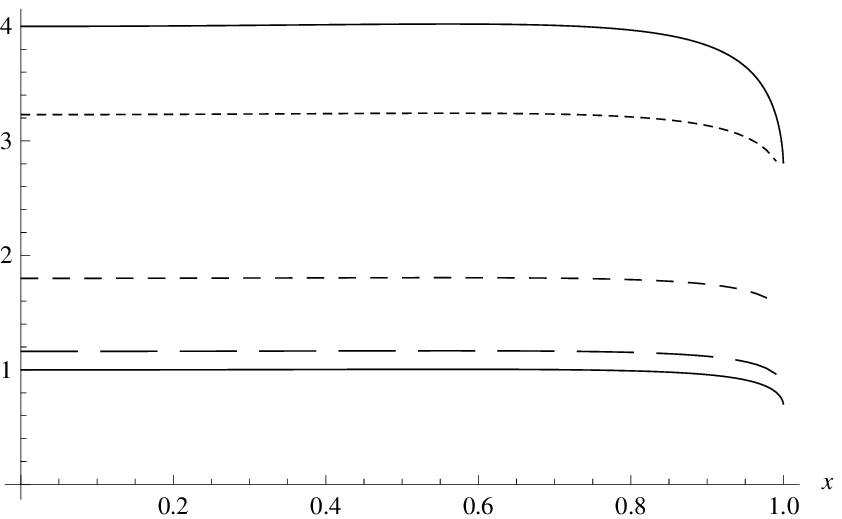}}}
\subfigure[]{\includegraphics[width=0.47\linewidth]{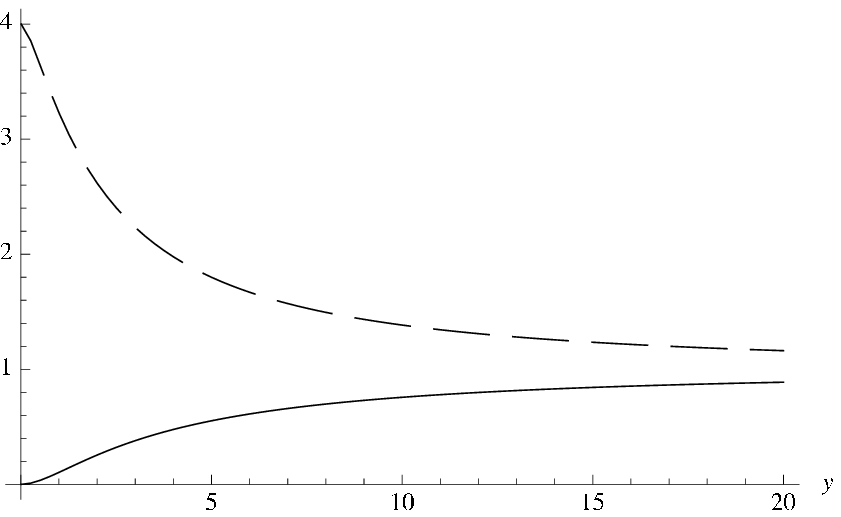}}
\caption{The suppression function $\mathcal{E}_1^2(x,y)$ for the first
  node (a) and $\mathcal{E}_2^2(x,y,2)$ for the second node (b) of the
  two nodes quiver is shown for various values of $y$, with the long,
  intermediate and short dashed lines corresponding to $y=20,5,1$,
  respectively.
  For $y\to\infty$, the MGM limit is recovered
  (solid line) for both (a) and (b), while for (b) the upper solid
  line corresponds to the second MGM limit $y\to 0$.
  In (c) the functions $\mathcal{E}_1^2(0,y)$ (solid line) and
  $\mathcal{E}_2^2(0,y,2)$ (dashed line) are shown. In this figure we
  chose $g_1=g_2$ and hence $\lambda_2 = 2$.}
\label{fig:2nodes}
\end{center}
\end{figure}

In fig.~\ref{fig:2nodes}, we display the functions
$\mathcal{E}_1^2(x,y)$ and $\mathcal{E}_2^2(x,y,\lambda_2)$ for various
values of $y$, for equal gauge couplings (and hence $\lambda_2 = 2$)
and we also present $\mathcal{E}_1^2(0,y),\mathcal{E}_2^2(0,y,2)$,
i.e.~the small $x$ regime interpolation between the above mentioned
MGM limits.

Finally, the suppression of the sfermion masses with respect to MGM is
$(\mathcal{E}_1^2)^{-1}:(\mathcal{E}_2^2)^{-1}=9.6:0.31$, for $y=1$,
equal gauge couplings and small $x$.
We see that when the matter and messenger are charged under
different nodes of the quiver, the suppression is relatively large,
even though we consider a ``hybrid gaugino-gauge mediation'' case,
where the messenger scale is equal to the mass of the heavy gauge particles,
$M=m_1$.

\subsection{Three nodes quivers}\label{sec:threenodes}

In this subsection we shall calculate the sfermion soft masses and the
link field soft masses for all three nodes quiver theories with a
single pair of messengers charged under one or two of the nodes.

\subsubsection{The basic three nodes quiver -- model q}\label{sec:modelP}

The fundamental quiver diagram is a triangle with a
single pair of messengers charged under one of the nodes, which we will
take to be $G_3$ (see fig.~\ref{fig:modelIIInodes1}a); the other cases
can be easily obtained from this one.
\begin{figure}[!tbp]
\begin{center}
\mbox{\subfigure[Model
    q]{\includegraphics[width=0.36\linewidth]{modelq}}\quad
  $\begin{subarray}{c}\mathop\Rightarrow \\ {v_{13}\to 0}\end{subarray}$ \quad
\subfigure[Linear quiver
  $N=3$]{\includegraphics[width=0.4\linewidth]{modelL3}}}\\
\mbox{\subfigure[Model
    p]{\includegraphics[width=0.36\linewidth]{modelp}}\quad
  $\begin{subarray}{c}\mathop\Rightarrow \\ {v_{13}\to 0}\end{subarray}$ \quad
\subfigure[Model T]{\includegraphics[width=0.4\linewidth]{modelT}}}
\caption{Quiver diagrams representing the theories with
the chiral matter superfields $Q_i$ charged under $G_i$
and link fields $L_{ij},\tilde{L}_{ij}$, connecting the $3$ gauge
groups. In (a,b) the messenger fields $T,\tilde{T}$ are charged only
under $G_{3}$, while in (c,d) they are only charged under $G_{2}$. (b)
is obtained from (a) by taking the limit $v_{13}\to 0$ and likewise
(d) is obtained from (c) in the same limit. }
\label{fig:modelIIInodes1}
\end{center}
\end{figure}
The mass-squared matrix for the gauge bosons is
\beq
\mathcal{M}_V^2 = 2
\begin{pmatrix}
g_1^2\left(v_{12}^2 + v_{13}^2\right) & - g_1 g_2 v_{12}^2 &
  - g_1 g_3 v_{13}^2 \\
- g_1 g_2 v_{12}^2 & g_2^2\left(v_{12}^2 + v_{23}^2\right) &
  - g_2 g_3 v_{23}^2 \\
- g_1 g_3 v_{13}^2 & - g_2 g_3 v_{23}^2 &
  g_3^2\left(v_{23}^2 + v_{13}^2\right)
\end{pmatrix} \ ,
\eeq
and it has the eigenvalues $0$ and
\begin{align}
m_{1,2}^2 &= A_{12}+A_{23}+A_{13} \mp \sqrt{(A_{12}+A_{23}+A_{13})^2
  - 4 P_2(\{g_i^2\})P_2(\{v_{ij}^2\})} \ , \label{eq:masses_modelq} \\
A_{ij} &\equiv \left(g_i^2 + g_j^2\right) v_{ij}^2 \ ,
\nonumber
\end{align}
where $P_2(\{x_i\}) \equiv x_1 x_2 + x_2 x_3 + x_1 x_3$.
After plugging the elements of the matrix $U$,
which diagonalize the above mass-squared matrix as in
eq.~\eqref{eq:Umatrix}, into eq.~\eqref{eq:formfactor},
one finds the following form factors for the matter $Q_1$, $Q_2$ and
$Q_3$, respectively,
\begin{align}
f_1^3(p^2) &= \left(
\frac{m_1^2m_2^2 - \zeta_{13} M^2 p^2}
{\left(p^2 - m_1^2\right)\left(p^2 - m_2^2\right)}
\right)^2 \ , \qquad
\zeta_{ij} M^2 \equiv \frac{2g_i^2g_j^2v_{ij}^2}{g_{\rm eff}^2} \ , \nonumber\\
f_2^3(p^2) &= \left(
\frac{m_1^2m_2^2 - \zeta_{23} M^2 p^2}
{\left(p^2 - m_1^2\right)\left(p^2 - m_2^2\right)}
\right)^2 \ , \\
f_3^3(p^2) &= \left(
\frac{m_1^2m_2^2 - \left(\zeta_{23} + \zeta_{13} +
  2\eta_{12}\lambda_3\right) M^2 p^2 + \lambda_3 p^4}
{\left(p^2 - m_1^2\right)\left(p^2 - m_2^2\right)}
\right)^2 \ , \qquad
\eta_{ij} M^2 \equiv A_{ij} \ , \qquad
\lambda_i \equiv \frac{g_i^2}{g_{\rm eff}^2} \ . \nonumber
\end{align}
The integrals \eqref{eq:sfermion_mass} have
a similar form for matter charged under any of the nodes of the various
quiver models, so it is convenient to define the following function
\begin{align}
&\mathcal{K}(x,y_1,y_2,z_1,z_2) \equiv
\frac{1}{x^2}\Bigg[
\alpha_0(x)
-\frac{2\left(y_1^2+y_2^2-z_1\right)}{y_1^2y_2^2} \beta_{-1}(x)
+\beta_0(x)
\nonumber
\end{align}
\begin{align}&
+\sum_{\begin{subarray}{c}i,j=1\\i\neq j\end{subarray}}^2
\frac{1}{(y_i^2-y_j^2)^2}\Bigg(
\frac{(y_j^2 - z_1 + y_i^2z_2)
\left(y_j^4 + (y_i^2+y_j^2)z_1 + y_i^4 z_2
- 3y_i^2y_j^2(1+z_2)\right)}{y_i^2-y_j^2} \alpha_1(x,y_i)
\non &
-y_i^2(y_j^2 - z_1 + y_i^2 z_2)^2 \alpha_2(x,y_i)
-\frac{2(y_j^2-z_1 + y_i^2 z_2)
\left(y_j^4 + y_i^2z_1 - y_i^2y_j^2(2+z_2)\right)}
{y_i^2(y_i^2-y_j^2)} \beta_1(x,y_i)
\non &
+(y_j^2-z_1 + y_i^2 z_2)^2 \beta_2(x,y_i)
\Bigg)
\Bigg] \ , \label{eq:Kfuncdef}
\end{align}
which corresponds to the generic form factor
\beq
f(p^2) = \left(
\frac{m_1^2m_2^2 - z_1 M^2 p^2 + z_2 p^4}
{\left(p^2 - m_1^2\right)\left(p^2 - m_2\right)^2}
\right)^2 \ .
\eeq
Now it is possible to see that the integrals \eqref{eq:sfermion_mass}
yield the following suppression functions
\begin{align}
\mathcal{E}_1^3(x,y_1,y_2,\zeta_{13}) &=
\mathcal{K}(x,y_1,y_2,\zeta_{13},0) \ , \non
\mathcal{E}_2^3(x,y_1,y_2,\zeta_{23}) &=
\mathcal{K}(x,y_1,y_2,\zeta_{23},0) \ , \\
\mathcal{E}_3^3(x,y_1,y_2,\zeta_{23},\zeta_{13},\eta_{12},\lambda_3) &=
\mathcal{K}(x,y_1,y_2,\zeta_{23}+\zeta_{13}+2\eta_{12}\lambda_3,\lambda_3)
\ , \nonumber
\end{align}
for the first, the second and the third node, respectively, of model q.

As explained in more detail in appendix \ref{app:limits}, the
functions $\alpha,\beta$ simplify in the limit $x\to 0$, which
is a good approximation for $x\lesssim 0.7$, where in turn the
function $\mathcal{K}$ simplifies as
\begin{align}
\mathcal{K}(0,y_1,y_2,z_1,z_2) &=
1-\frac{2\left(y_1^2+y_2^2-z_1\right)}{y_1^2y_2^2}
+\sum_{\begin{subarray}{c}i,j=1\\i\neq j\end{subarray}}^2
\frac{1}{(y_i^2-y_j^2)^2}\Bigg(
(y_j^2-z_1 + y_i^2 z_2)^2 \tilde{\beta}_2(y_i)
\non & \phantom{=\ }
-\frac{2(y_j^2-z_1 + y_i^2 z_2)
\left(y_j^4 + y_i^2z_1 - y_i^2y_j^2(2+z_2)\right)}
{y_i^2(y_i^2-y_j^2)} \tilde{\beta}_1(y_i)
\Bigg) \ ,
\end{align}
with $\tilde{\beta}_{1,2}(y) = \lim_{x\to 0}\beta_{1,2}(x,y)/x^2$
being the limits given in eqs.~\eqref{eq:x0limitbeta1} and
\eqref{eq:x0limitbeta2}.

Finally, the suppression of the sfermion masses with respect to MGM is
$(\mathcal{E}_1^3)^{-1}:(\mathcal{E}_2^3)^{-1}:(\mathcal{E}_3^3)^{-1}
=9.6:9.6:0.15$,
for $y_1=1$, equal gauge couplings, equal VEVs and small $x$.

\subsubsection{Model p}

Although this model is equivalent to the previous one --
the two are related by moving the messengers $T,\tilde{T}$
from the node $G_3$ to $G_2$,
as depicted in fig.~\ref{fig:modelIIInodes1}c --
we introduce it to simplify the discussion of other, non-equivalent models.
By symmetry, the result for the suppression functions is
\begin{align}
\mathcal{E}_1^2(x,y_1,y_2,\zeta_{12}) &=
\mathcal{K}(x,y_1,y_2,\zeta_{12},0) \ , \non
\mathcal{E}_2^2(x,y_1,y_2,\zeta_{12},\zeta_{23},\eta_{13},\lambda_2) &=
\mathcal{K}(x,y_1,y_2,\zeta_{12}+\zeta_{23}+2\eta_{13}\lambda_2,\lambda_2)
\ , \\
\mathcal{E}_3^2(x,y_1,y_2,\zeta_{23}) &=
\mathcal{K}(x,y_1,y_2,\zeta_{23},0) \ . \nonumber
\end{align}

\subsubsection{The linear quiver $N=3$}

Using model q, one obtains the suppression functions for the
linear quiver with $N=3$ nodes, as shown in
fig.~\ref{fig:modelIIInodes1}b, by taking the limit $v_{13}\to 0$.
The masses $m_\ell$ of eq.~\eqref{eq:masses_modelq} become
their respective limit for vanishing $v_{13}$. Hence, the result for
the suppression functions corresponding to matter charged under any of
the three nodes of the linear quiver reads
\begin{align}
\mathcal{E}_1^3(x,y_1,y_2) &=
\mathcal{K}(x,y_1,y_2,0,0) \ , \non
\mathcal{E}_2^3(x,y_1,y_2,\zeta_{23}) &=
\mathcal{K}(x,y_1,y_2,\zeta_{23},0) \ , \\
\mathcal{E}_3^3(x,y_1,y_2,\zeta_{23},\eta_{12},\lambda_3) &=
\mathcal{K}(x,y_1,y_2,\zeta_{23}+2\eta_{12}\lambda_3,\lambda_3)
\ . \nonumber
\end{align}

\begin{figure}[tbp]
\begin{center}
\mbox{\subfigure[]{\includegraphics[width=0.47\linewidth]{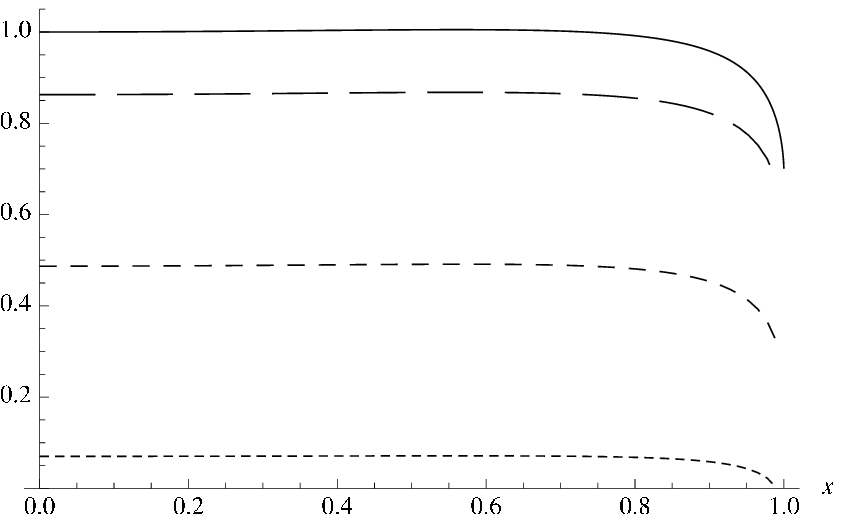}}\quad
\subfigure[]{\includegraphics[width=0.47\linewidth]{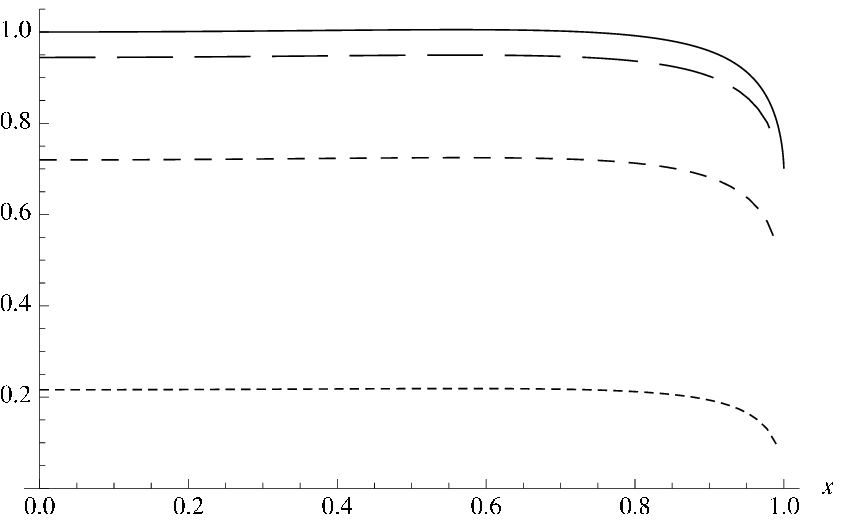}}}
\mbox{\subfigure[]{\includegraphics[width=0.47\linewidth]{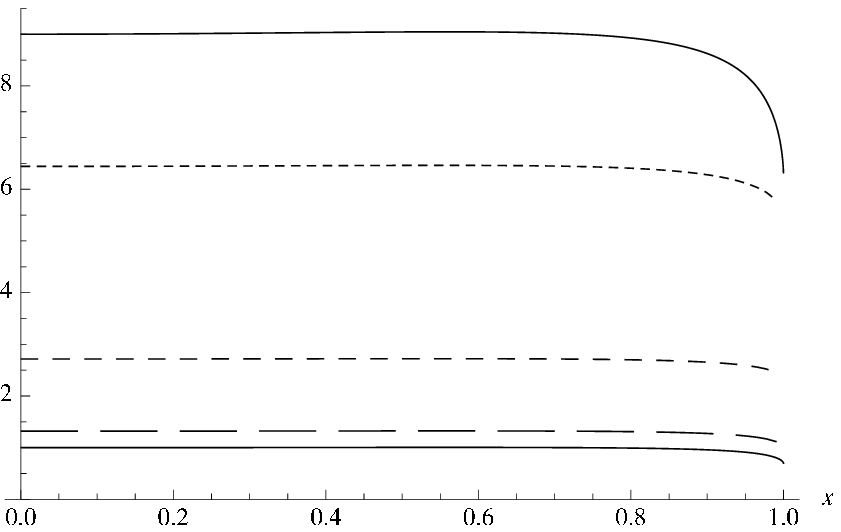}}\quad
\subfigure[]{\includegraphics[width=0.47\linewidth]{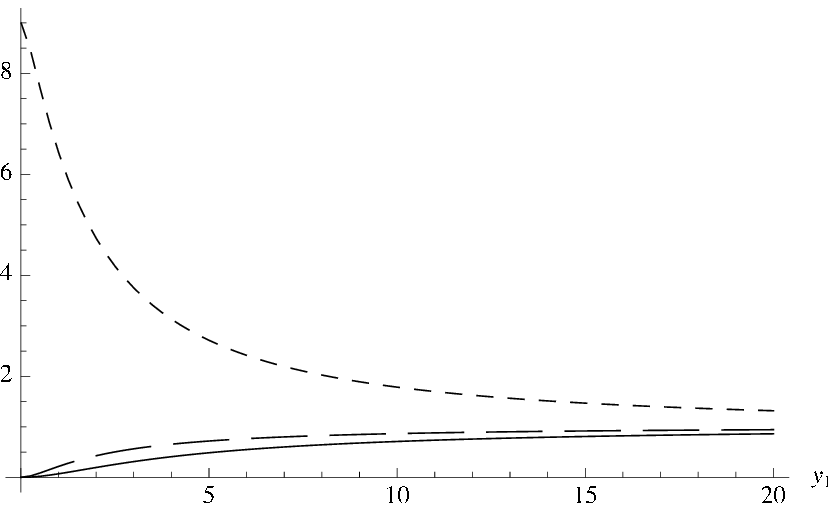}}}
\caption{The suppression functions (a) $\mathcal{E}_1^3(x,y_1,y_2)$
  for the first node, (b) $\mathcal{E}_2^3(x,y_1,y_2,y_2^2)$ for the
  second node and (c) $\mathcal{E}_3^3(x,y_1,y_2,y_2^2,y_1^2,3)$ for
  the third node of the linear three node quiver for various values of
  $y_1$, with the long, intermediate and short dashed lines
  corresponding to $y_1=20,5,1$, respectively.
  For the first two nodes, the MGM limit is
  recovered (solid line) for $y \rightarrow \infty$, while for the
  third node the two solid lines correspond to the two MGM limits,
  $y\to\infty$ and $y\to 0$.
  Finally, (d) depicts all the functions $\mathcal{E}_{1,2,3}^3$ as
  function of $y_1$ for small $x$ with solid, long dashed and
  short dashed lines, respectively.
  In this figure, we have chosen $g_1=g_2=g_3$, $v_{12}=v_{23}$ and
  hence $y_2=\sqrt{3} y_1$, $\zeta = y_2^2$, $\lambda_1 = 3y_2^2$ and
  $\lambda_2 = 3$. }
\label{fig:linear3nodes}
\end{center}
\end{figure}

In fig.~\ref{fig:linear3nodes}, we display the functions
$\mathcal{E}_{1,2,3}^3$ for various values of $y_1$, for equal VEVs and
equal gauge couplings (and hence $y_2=\sqrt{3} y_1$,
$\zeta_{23} = y_2^2$, $\eta_{12} = y_2^2/3$ and $\lambda_3 = 3$) and
we also present $\mathcal{E}_{1,2,3}^3$ in the small $x$ regime
as functions of $y_1$.
In the limit of $y_{1,2}\to\infty$ (i.e.~taking
$v_{12},v_{23}\to\infty$), the function
$\mathcal{K}$ of eq.~\eqref{eq:Kfuncdef} reveals that, for each node,
the suppression function reduces to the MGM one \eqref{eq:Efunc_MGM}.
In the opposite limit, $y_{1,2}\to 0$ (i.e.~taking $v_{12},v_{23}\to 0$),
the suppression function goes to zero for matter charged under
the first and second nodes,
while it gives the MGM result with a factor of $\lambda_3^2$
for matter charged under the third node $G_3$,
\beq
\mathcal{E}_3^3(x,0,0,0,\lambda_3) = \frac{\lambda_3^2}{x^2}
\left(\alpha_0(x) + \beta_0(x)\right) \ . \
\eeq

Finally, the suppression of the sfermion masses with respect to MGM is
$(\mathcal{E}_1^3)^{-1}:(\mathcal{E}_2^3)^{-1}:(\mathcal{E}_3^3)^{-1}
=14.3:4.6:0.16$, for $y_1=1$, equal gauge couplings, equal VEVs and small
$x$. In this case, the suppression is somewhat larger relative to the
previous examples, for matter charged under $G_1$.

\subsubsection{Model T}

Similarly, using model p, it is easy obtain the suppression functions for model
T shown in fig.~\ref{fig:modelIIInodes1}d, by taking the limit
$v_{13}\to 0$. The masses $m_\ell$ become their respective limit of
vanishing $v_{13}$. Hence, the result for the suppression functions reads
\begin{align}
\mathcal{E}_1^2(x,y_1,y_2,\zeta_{12}) &=
\mathcal{K}(x,y_1,y_2,\zeta_{12},0) \ , \non
\mathcal{E}_2^2(x,y_1,y_2,\zeta_{12},\zeta_{23},\lambda_2) &=
\mathcal{K}(x,y_1,y_2,\zeta_{12}+\zeta_{23},\lambda_2) \ , \\
\mathcal{E}_3^2(x,y_1,y_2,\zeta_{23}) &=
\mathcal{K}(x,y_1,y_2,\zeta_{23},0) \ . \nonumber
\end{align}
The ratio of suppression with respect to MGM is
$(\mathcal{E}_1^2)^{-1}:(\mathcal{E}_2^2)^{-1}:(\mathcal{E}_3^2)^{-1}
=4.6:0.18:4.6$,
for $y_1=1$, equal gauge couplings, equal VEVs and small $x$.

\subsubsection{Bifundamental messenger models}

\begin{figure}[!tbp]
\begin{center}
\mbox{\subfigure[Model
    p+q]{\includegraphics[width=0.36\linewidth]{modeldelta}}\quad
\subfigure[Model F]{\includegraphics[width=0.38\linewidth]{modelII}}}\\
\mbox{\subfigure[Model
    $\Lambda$]{\includegraphics[width=0.36\linewidth]{modellambda}}}
\caption{Quiver diagrams representing the theories with
the chiral matter superfields $Q_i$ charged under $G_i$
and link fields $L_{ij},\tilde{L}_{ij}$, connecting the $3$ gauge
groups. The messenger fields $T (\tilde{T})$ are (anti-)bifundamentals
of $G_2\times G_3$. }
\label{fig:3nodes2messengers}
\end{center}
\end{figure}

Here we consider models in which the messenger fields are charged under
two of the gauge groups, $G_2$ and $G_3$.
To compute the sfermion masses in the p+q model
(fig.~\ref{fig:3nodes2messengers}a),
using eq.~\eqref{eq:sparticle_masses},
all we need is to superpose the suppression functions of models q and p,
weighted by the Dynkin indices $n(\mathbf{s}_2)$, $n(\mathbf{s}_3)$ of
the messenger pair on $G_{2,3}$, respectively; one thus obtains the
factors
\beq
n(\mathbf{s}_2) \mathcal{E}_i^2 + n(\mathbf{s}_3) \mathcal{E}_i^3
\ . \label{eq:massfunction2messengers}
\eeq
For instance, when both groups are $SU(r)$, the suppression
factors of model p+q are $r(\mathcal{E}_i^2 + \mathcal{E}_i^3)$.

{}For model F (fig.~\ref{fig:3nodes2messengers}b), the suppression factors can
be obtained either as the limit $v_{13}\to 0$ of model p+q or, equivalently,
as the sum of the linear quiver $N=3$ and model T factors, as given in
eq.~\eqref{eq:massfunction2messengers}.
Finally, model $\Lambda$ can be obtained e.g.~from model p+q by taking
the limit $v_{23}\to 0$.

\subsection{More nodes}\label{sec:morenodes}

\begin{figure}[!tbp]
\begin{center}
\includegraphics[width=0.8\linewidth]{quiver3}
\caption{A quiver diagram representing a class of quiver theories with
the chiral matter superfields $Q_i$ charged under $G_i$
and $N-1$ link fields $L_{i,i+1},\tilde{L}_{i,i+1}$, $i=1,\ldots,N-1$,
connecting the $N$ gauge groups and, finally, the messenger fields
$T,\tilde{T}$ are charged only under the last group $G_N$. We call
this class of models linear quivers. }
\label{fig:linear_quiver}
\end{center}
\end{figure}

Linear quivers, presented in fig.~\ref{fig:linear_quiver}, have
a particularly simple structure, which we consider in this subsection.
For equal gauge couplings $g$ and equal VEVs $v$, the form factors
were computed for arbitrary $N$ in \cite{McGarrie}.
The vector boson mass-squared matrix takes the form
\beq
\mathcal{M}^2_V = 2 g^2 v^2 \begin{pmatrix}
1 & -1 & & &  \\
-1 & 2 & -1 & &  \\
 & \ddots & \ddots & \ddots &  \\
 & & -1 & 2 & -1   \\
  & & & -1 & 1 \\
 \end{pmatrix} \ .
\eeq
The eigenvalues and the diagonalizing matrix $U$, respectively, are
\cite{Hill:2000mu,Cheng:2001vd,Csaki:2001em}:
\beq
m_k^2 = 8 g^2 v^2 \sin^2 \left( \frac{(k-1) \pi}{2 N} \right) \ , \qquad
U_{i j}= \left(\frac{2}{2^{\delta_{i1}} N} \right)^{1/2}
\cos \frac{(i-1) (2 j-1) \pi}{2 N} \ ,
\label{tantilinks}
\eeq
where $1 \leq k \leq N$ (in this formula a slightly different convention
is used: $m_1=0$, $m_k \neq 0$ for $k>1$).
Using eq.~(\ref{tantilinks}) in eq.~(\ref{eq:formfactor}) we recover
the result of \cite{McGarrie}.
In the limit of large $N$, the extra dimension setting
is recovered (see
e.g.~\cite{Mirabelli:1997aj,Kaplan:1999ac,Chacko:1999mi,McGarrieRusso}).
In the case of general VEVs and couplings such a compact
analytic expression does not materialize and one has to diagonalize
a generic tridiagonal $N \times N$ matrix in order to find
the form factors.

We will consider in detail some properties in two examples -- the linear
quivers of fig.~\ref{fig:linear_quiver} with $N=4,5$.
In the $N=4$ case, the mass-squared matrix for the gauge bosons is
given by
\beq
\mathcal{M}_V^2 = 2
\begin{pmatrix}
g_1^2 v_{12}^2 & -g_1 g_2 v_{12}^2 & 0 & 0 \\
-g_1 g_2 v_{12}^2 & g_2^2 \left(v_{12}^2 + v_{23}^2\right) & -g_2 g_3 v_{23}^2 & 0 \\
0 & -g_2 g_3 v_{23}^2 & g_3^2 \left(v_{23}^2 + v_{34}^2\right) & -g_3 g_4 v_{34}^2
\\
0 & 0 & -g_3 g_4 v_{34}^2 & g_4^2 v_{34}^2
\end{pmatrix} \ , \label{eq:N4massmatrix}
\eeq
and the eigenvalues are in general quite complicated.
The generic form factor for matter charged under the first node $G_1$ is
\beq
f_1^4(p^2) = \left(
\frac{m_1^2m_2^2m_3^2}{\left(p^2-m_1^2\right)\left(p^2-m_2^2\right)
\left(p^2-m_3^2\right)}\right)^2 \ ,
\eeq
giving rise to the suppression function
\begin{align}
\mathcal{E}_1^4(x,y_1,y_2,y_3) = \frac{1}{x^2}\Bigg[&\alpha_0(x)
-2\frac{y_1^2y_2^2 + y_2^2y_3^2 + y_1^2y_3^2}{y_1^2y_2^2y_3^2}
\beta_{-1}(x) + \beta_0(x) \non &
-\sum_{\begin{subarray}{c}i,j,k \\ {\rm cyclic}\end{subarray}}\Bigg(
\frac{y_j^4y_k^4[5y_i^4 - 3(y_j^2 + y_k^2)y_i^2 + y_j^2y_k^2]}
{(y_i^2 - y_j^2)^3(y_i^2 - y_k^2)^3}
\alpha_1(x,y_i)
\non & \phantom{-\sum_{\begin{subarray}{c}i,j,k \\ {\rm
        cyclic}\end{subarray}}\Bigg(\ }
+\frac{y_j^4y_k^4}
{(y_i^2 - y_j^2)^2(y_i^2 - y_k^2)^2} \left(y_i^2\alpha_2(x,y_i)
+\beta_2(x,y_i)\right)
\non & \phantom{-\sum_{\begin{subarray}{c}i,j,k \\ {\rm
        cyclic}\end{subarray}}\Bigg(\ }
+\frac{y_j^4y_k^4[6y_i^4 - 4(y_j^2 + y_k^2)y_i^2 + y_j^2y_k^2]}
{y_i^2(y_i^2 - y_j^2)^3(y_i^2 - y_k^2)^3}
\beta_1(x,y_i)
\Bigg)\Bigg] \ .
\end{align}
Let us stress that this suppression function is valid for any
$g_i$, $v_{ij}$.

{}For the remaining nodes, for simplicity, we present just the case of
equal couplings and equal VEVs.
The form factors can be computed straightforwardly from eqs.~(\ref{eq:formfactor}) and
(\ref{tantilinks}). The corresponding suppression functions $\mathcal{E}_i^4(x,y_1)$
can be expressed in the form of eq.~(\ref{eq:Efunc_def}).
 The resulting coefficients of the
suppression functions are given in table \ref{tab:L4coefficients}.
The suppression of the sfermion masses with respect to MGM is
$(\mathcal{E}_1^4)^{-1}:(\mathcal{E}_2^4)^{-1}:(\mathcal{E}_3^4)^{-1}
: (\mathcal{E}_4^4)^{-1}
=16.2:9.3:2.3:0.098$,
for $y_1=1$, equal gauge couplings, equal VEVs and small $x$.

\begin{sidewaystable}[!p]
\centering
\begin{tabular}{c||ccccccccccccccc}
 & $a_0$ & $a_{1,1}$ & $a_{1,2}$ & $a_{1,3}$ &
  $\frac{a_{2,1}}{M^2}$ & $\frac{a_{2,2}}{M^2}$ &
  $\frac{a_{2,3}}{M^2}$ & $b_{-1}M^2$ & $b_0$ & $b_{1,1}M^2$ &
  $b_{1,2}M^2$ & $b_{1,3}M^2$ & $b_{2,1}$ & $b_{2,2}$ & $b_{2,3}$ \\
\hline
\hline
$\mathcal{E}_1^4$ & $1$ & $\frac{1}{\sqrt{2}}$ & $-1$ &
$-\frac{1}{\sqrt{2}}$ & $-\frac{1}{2}y_3^2$ &
$-y_2^2$ & $-\frac{1}{2}y_1^2$ & $-\frac{10}{y_2^2}$ & $1$ & $\frac{8+5
  \sqrt{2}}{2 y_2^2}$ & $\frac{2}{y_2^2}$ & $\frac{8-5 \sqrt{2}}{2
  y_2^2}$ & $\frac{3}{2}+\sqrt{2}$ & $1$ & $\frac{3}{2}-\sqrt{2}$ \\
$\mathcal{E}_2^4$ & $1$ & $-2+\frac{1}{\sqrt{2}}$ & $3$ &
$-2-\frac{1}{\sqrt{2}}$ & $-\frac{1}{2}y_1^2$ &$-y_2^2$ &
$-\frac{1}{2}y_3^2$ & $-\frac{6}{y_2^2}$ & $1$ & $\frac{8+3\sqrt{2}}{2
  y_2^2}$ & $-\frac{2}{y_2^2}$ & $\frac{8-3\sqrt{2}}{2 y_2^2}$ &
$\frac{1}{2}$ & $1$ & $\frac{1}{2}$ \\
$\mathcal{E}_3^4$ & $1$ & $2+\frac{1}{\sqrt{2}}$ & $-5$ &
$2-\frac{1}{\sqrt{2}}$ & $-\frac{1}{2}y_1^2$ & $-y_2^2$ &
$-\frac{1}{2}y_3^2$ & $\frac{2}{y_2^2}$ & $1$ & $-\frac{8+5
  \sqrt{2}}{2 y_2^2}$ & $\frac{6}{y_2^2}$ & $-\frac{8-5 \sqrt{2}}{2
  y_2^2}$ & $\frac{1}{2}$ & $1$ & $\frac{1}{2}$ \\
$\mathcal{E}_4^4$ & $1$ & $4+\frac{1}{\sqrt{2}}$ & $7$ &
$4-\frac{1}{\sqrt{2}}$ & $-\frac{1}{2}y_3^2$ & $-y_2^2$ &
$-\frac{1}{2}y_1^2$ & $\frac{14}{y_2^2}$ & $1$ &
$-\frac{8+3\sqrt{2}}{2 y_2^2}$ & $-\frac{6}{y_2^2}$ &
$-\frac{8-3\sqrt{2}}{2 y_2^2}$ & $\frac{3}{2}+\sqrt{2}$ & $1$ &
$\frac{3}{2}-\sqrt{2}$
\end{tabular}
\caption{Coefficients in the suppression functions for the linear quiver with
  $N=4$, shown in fig.~\ref{fig:linear_quiver}, with equal couplings and
  equal VEVs. The suppression functions are given in
  eq.~\eqref{eq:Efunc_def}. }
\label{tab:L4coefficients}

\bigskip

\begin{tabular}{c||ccccccccc}
 & $a_0$ & $a_{1,1}$ & $a_{1,2}$ & $a_{1,3}$ & $a_{1,4}$ &
  $\frac{a_{2,1}}{M^2}$ & $\frac{a_{2,2}}{M^2}$ &
  $\frac{a_{2,3}}{M^2}$ & $\frac{a_{2,4}}{M^2}$ \\
\hline
\hline
$\mathcal{E}_1^5$ &
$1$ &
$\frac{\sqrt{5}}{2}$ &
$-\frac{1}{2}$ &
$-\frac{\sqrt{5}}{2}$ & $-\frac{1}{2}$ &
$-\frac{5\left(3+\sqrt{5}\right)y_1^2}{8}$ &
$-\frac{\left(25+11\sqrt{5}\right) y_1^2}{8}$ &
$-\frac{5\left(3+\sqrt{5}\right)y_1^2}{8}$ &
$-\frac{\left(5-\sqrt{5}\right)y_1^2}{8}$ \\
$\mathcal{E}_2^5$ & $1$ & $-\frac{\sqrt{5}}{2}$ &
$-\frac{1}{2}+\sqrt{5}$ &
$\frac{\sqrt{5}}{2}$ &
$-\frac{1}{2}-\sqrt{5}$ & $-\frac{5y_1^2}{4}$ &
$-\frac{\left(5+\sqrt{5}\right)y_1^2}{8}$ &
$-\frac{5\left(7+3 \sqrt{5}\right)y_1^2}{8}$ &
$-\frac{\left(5+2 \sqrt{5}\right)y_1^2}{4}$ \\
$\mathcal{E}_3^5$ & $1$ & $0$ &
$-\frac{1-\sqrt{5}}{2}$ & $0$ &
$-\frac{1+\sqrt{5}}{2}$ & $0$ &
$-\left(5+2 \sqrt{5}\right)y_1^2$ & $0$ &
$-\frac{\left(5+\sqrt{5}\right)y_1^2}{2}$ \\
$\mathcal{E}_4^5$ & $1$ & $2 \sqrt{5}$ &
$-\frac{1+3\sqrt{5}}{2}$ & $-2 \sqrt{5}$ &
$-\frac{1-3 \sqrt{5}}{2}$ &
$-\frac{5y_1^2}{4}$ &
$-\frac{\left(5+\sqrt{5}\right)y_1^2}{8}$ &
$-\frac{5\left(7+3\sqrt{5}\right)y_1^2}{8}$ &
$-\frac{\left(5+2\sqrt{5}\right)y_1^2}{4}$ \\
$\mathcal{E}_5^5$ & $1$ & $\frac{25-3\sqrt{5}}{4}$ &
$\frac{23+5 \sqrt{5}}{4}$ &
$\frac{25+3 \sqrt{5}}{4}$ &
$\frac{23-5 \sqrt{5}}{4}$ &
$-\frac{5\left(3+\sqrt{5}\right)y_1^2}{8}$ &
$-\frac{\left(25+11 \sqrt{5}\right)y_1^2}{8}$ &
$-\frac{5\left(3+\sqrt{5}\right)y_1^2}{8}$ &
$-\frac{\left(5-\sqrt{5}\right)y_1^2}{8}$
\end{tabular}
\caption{Coefficients of $\alpha$ type in the suppression functions for the
  linear quiver with $N=5$, shown in fig.~\ref{fig:linear_quiver}, with
  equal couplings and equal VEVs. The suppression functions are given in
  eq.~\eqref{eq:Efunc_def}. }
\label{tab:L5coefficients1}

\bigskip

\begin{tabular}{c||cccccccccc}
 & $b_{-1}M^2$ & $b_0$ &
  $b_{1,1}M^2$ & $b_{1,2}M^2$ & $b_{1,3}M^2$ & $b_{1,4}M^2$ &
  $b_{2,1}$ & $b_{2,2}$ & $b_{2,3}$ & $b_{2,4}$ \\
\hline
\hline
$\mathcal{E}_1^5$ & $-\frac{4 \left(3-\sqrt{5}\right)}{y_1^2}$ &
$1$ & $\frac{15+\sqrt{5}}{8y_1^2}$ &
$\frac{10+\sqrt{5}}{20y_1^2}$ & $\frac{30-13 \sqrt{5}}{4y_1^2}$ &
$\frac{85-37\sqrt{5}}{40y_1^2}$ &
$\frac{5\left(3+\sqrt{5}\right)}{8}$ &
$\frac{7+3\sqrt{5}}{8}$ &
$\frac{5\left(3-\sqrt{5}\right)}{8}$ &
$\frac{7-3 \sqrt{5}}{8}$ \\
$\mathcal{E}_2^5$ & $-\frac{3\left(3-\sqrt{5}\right)}{y_1^2}$ &
$1$ &
$\frac{5+2\sqrt{5}}{4y_1^2}$ &
$\frac{35-23 \sqrt{5}}{40y_1^2}$ &
$\frac{65-29 \sqrt{5}}{8y_1^2}$ &
$-\frac{25-14 \sqrt{5}}{20y_1^2}$ &
$\frac{5}{4}$ & $\frac{1}{4}$ & $\frac{5}{4}$ &
$\frac{1}{4}$ \\
$\mathcal{E}_3^5$ & $-\frac{3-\sqrt{5}}{y_1^2}$ & $1$ &
$0$ & $\frac{5-\sqrt{5}}{5y_1^2}$ & $0$ &
$\frac{2-\frac{4}{\sqrt{5}}}{y_1^2}$ &
$0$ & $\frac{3+\sqrt{5}}{2}$ & $0$ &
$\frac{3-\sqrt{5}}{2}$ \\
$\mathcal{E}_4^5$ & $\frac{6-2 \sqrt{5}}{y_1^2}$ & $1$ &
$\frac{5-8 \sqrt{5}}{4y_1^2}$ &
$-\frac{3\left(5-9\sqrt{5}\right)}{40y_1^2}$ &
$-\frac{85-41\sqrt{5}}{8y_1^2}$ &
$\frac{75-36\sqrt{5}}{20y_1^2}$ &
$\frac{5}{4}$ &
$\frac{1}{4}$ & $\frac{5}{4}$ & $\frac{1}{4}$ \\
$\mathcal{E}_5^5$ & $\frac{6\left(3-\sqrt{5}\right)}{y_1^2}$ &
$1$ & $-\frac{35-11\sqrt{5}}{8y_1^2}$ &
$-\frac{40-\sqrt{5}}{20y_1^2}$ &
$-\frac{20-7\sqrt{5}}{4y_1^2}$ &
$-\frac{265-113\sqrt{5}}{40y_1^2}$ &
$\frac{5\left(3+\sqrt{5}\right)}{8}$ &
$\frac{7+3\sqrt{5}}{8}$ &
$\frac{5\left(3-\sqrt{5}\right)}{8}$ &
$\frac{7-3\sqrt{5}}{8}$
\end{tabular}
\caption{Coefficients of $\beta$ type in the suppression functions for the
  linear quiver with $N=5$, shown in fig.~\ref{fig:linear_quiver}, with
  equal couplings and equal VEVs. The suppression functions are given in
  eq.~\eqref{eq:Efunc_def}. }
\label{tab:L5coefficients2}

\end{sidewaystable}

For the linear quiver of fig.~\ref{fig:linear_quiver} with $N=5$, the
mass-squared matrix for the gauge bosons is given by
\beq
\mathcal{M}_V^2 = 2
\begin{pmatrix}
g_1^2 v_{12}^2 & -g_1 g_2 v_{12}^2 & 0 & 0 & 0\\
-g_1 g_2 v_{12}^2 & g_2^2 \left(v_{12}^2 + v_{23}^2\right) & -g_2 g_3
v_{23}^2 & 0 & 0\\
0 & -g_2 g_3 v_{23}^2 & g_3^2 \left(v_{23}^2 + v_{34}^2\right) & -g_3
g_4 v_{34}^2 & 0\\
0 & 0 & -g_3 g_4 v_{34}^2 & g_4^2 \left(v_{34}^2 + v_{45}^2\right) &
-g_4 g_5 v_{45}^2 \\
0 & 0 & 0 & -g_4 g_5 v_{45}^2 & g_5^2 v_{45}^2
\end{pmatrix} \ , \label{eq:N5massmatrix}
\eeq
and again the eigenvalues are in general quite complicated.
The generic form factor for matter charged under the first node $G_1$ is
\beq
f_1^5(p^2) = \left(\frac{m_1^2m_2^2m_3^2m_4^2}
{\left(p^2-m_1^2\right)\left(p^2-m_2^2\right)
\left(p^2-m_3^2\right)\left(p^2-m_4^2\right)}\right)^2 \ .
\eeq
The form factors for the other nodes can be computed
as in the previous example. For equal $g_i$ and $v_{ij}$,
the resulting coefficients of the suppression functions
are given in tables \ref{tab:L5coefficients1},
 \ref{tab:L5coefficients2}.
 The suppression of the sfermion masses with respect to MGM is
$(\mathcal{E}_1^5)^{-1}:(\mathcal{E}_2^5)^{-1}:(\mathcal{E}_3^5)^{-1}
:(\mathcal{E}_4^5)^{-1}:(\mathcal{E}_5^5)^{-1}
=17.2:12.4:5.6:1.3:0.069$,
for $y_1=1$, equal gauge couplings, equal VEVs and small $x$.

\begin{figure}[!htp]
\begin{center}
\includegraphics[width=0.4\linewidth]{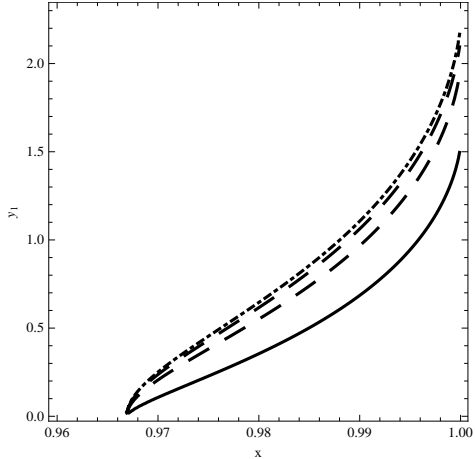}
\caption{Contour plot illustrating the points in the ($x,y_1$)-plane
  where the suppression function $\mathcal{E}_1^N$ vanishes,
  for $N=2,3,4,5$. The $N=2$ line is the lowest one while for each
  increasing $N$ the line moves slightly upwards in the plot. We have
  taken all couplings equal $g_i=g$ and all VEVs equal $v_{ij}=v$ in
  this plot. The tachyonic regime in each case is below its corresponding contour.}
\label{fig:tachreg}
\end{center}
\end{figure}
Finally, let us discuss a feature of the sfermion masses, found in
\cite{Auzzi:2010mb}, viz.~they become tachyonic
in the large $x\lesssim 1$ regime. We consider here the
tachyonic regime in the case of the linear quiver
(fig.~\ref{fig:linear_quiver}) for a sfermion charged under $G_1$,
whose suppression function is $\mathcal{E}_1^N$, as the number of nodes $N$ increases.
The result for $N=2,3,4,5$ is presented in fig.~\ref{fig:tachreg};
one can see that the tachyonic regime increases with $N$,
but only to a certain extent.

\section{Discussion}\label{sec:discussion}

In this work,
we computed the two-loop contributions to the scalar
soft masses in supersymmetric quiver gauge theories
with a general matter content.
Our results are valid in the hybrid regime --
when the messenger scale $M$ is comparable to the
masses $m_\ell$ of the additional gauge particles.
On the other hand,
three-loop contributions become important when $m_\ell/M$ is small
(see e.g.~eq.~(5.9) in \cite{Auzzi:2011gh} for a recent evaluation
of the relative 3-loops/2-loops contribution in some cases).
The suppression of sfermion masses is significant
-- even in the hybrid regime --
when the matter and messengers of SUSY breaking
are not charged under the same group.

In phenomenological applications,
the unbroken gauge group of the quiver gauge theory
must be the SM one at low energy.
This can be achieved, e.g., by taking $G_i=SU(5)$
in each node with, say, only
$G_1$ being broken to $SU(3)\times SU(2)\times U(1)$.
In this example, unification is manifest (if the Higgses are charged only under $G_1$).
Moreover, such a grand unified theory may be perturbative
if the number of nodes is sufficiently small
(up to about five nodes).

To find the physical pole masses,
one should consider, of course, the RGE from
the messenger scale down to the weak scale.
This can be done straightforwardly,
as was done recently for some two nodes examples
in \cite{Auzzi:2010xc,Auzzi:2011gh}.
We have done it in additional examples and found,
in particular, that the rather large suppression
of the scalar masses gives rise to exotic sparticle spectra,
even in the hybrid case.
In particular,
similar to the results in \cite{Auzzi:2010xc,Auzzi:2011gh},
the stop mass can be sufficiently small
-- to provide an explanation to the hierarchy problem --
even in models where the gaugino masses
vanish to leading order in SUSY breaking
(which is a typical property of quiver models
that have a dynamical embedding
in some higher energy theory \cite{Green:2010ww}),
and both the right-handed as well as the left-handed sleptons
can be lighter than the bino in the low-scale mediation regime
(when the messenger scale $M$ is comparable
to the effective SUSY-breaking scale $F/M$).

It is interesting to mention that, in the limit of small $m_{\ell}/M$,
the Higgs mass gets a contribution which is absent in the MSSM
\cite{Batra:2003nj, Maloney:2004rc};
this comes from the D-terms of the heavy gauge bosons,
which do not decouple completely in presence of SUSY breaking.
In some parts of the parameter space this can
raise the Higgs mass above 114 GeV in a way which is
compatible with naturalness constraints; 
see e.g.~the recent works \cite{Craig:2011yk,Auzzi:2011gh}.
{}For the mechanism to be effective, $m_{\ell}$
should be of the order of a few TeV. Furthermore,
the part of parameter space in which this effect is not negligible
is not compatible with the usual gauge coupling unification;
it could be compatible with  accelerated unification
 \cite{ArkaniHamed:2001vr}, instead.

Quiver gauge theories
that have dynamical embedding in (deformed) SQCD
are of particular interest.
In this paper, for simplicity, we focused on models
with a minimal messenger sector.
The generalization of our analysis to any weakly coupled
messengers sector -- in particular, to those realized in
the dynamical embedding of \cite{Green:2010ww} --
is straightforward, and can be done as in \cite{Auzzi:2011gh}.

The analysis of this work lays the grounds for further
investigation of the various constraints which arise in
``(de)constructing a natural and flavorful
supersymmetric standard model,'' as in \cite{Craig:2011yk}.
Addressing the texture of the Yukawa couplings
and the masses of the higgsinos and Higgs particles
-- hopefully in an appropriate way to ameliorate
the flavor puzzle and the $\mu/B\mu$ problem --
can be done by separating the three generations
and Higgs superfields on different nodes of the quiver.
Consequently,
the smallness of parameters in the Yukawa matrices
and the Higgs mass terms is natural, since they arise
from higher dimension operators in the effective action
(being suppressed e.g.~by the SQCD scale of
the high-energy embedding theory discussed above).
Moreover, inverted sparticle hierarchies
-- which are less constrained by current LHC limits --
are obtained when the first two generations and the
messengers are charged under the same nodes.
The detailed investigation of this interesting application
is left for future work.

Finally,
in this note we have limited our analysis
to quiver models with fields in the bifundamental
plus anti-bifundamental representation
in each of the links.
A phenomenologically appealing extension
of our investigation is to consider more generic
representations for the link fields;
see e.g.~the recent work \cite{Barbieri:2011ci}.

\subsubsection*{Acknowledgments}

We thank Zohar Komargodski for discussions.
This work was supported in part
by the BSF -- American-Israel Bi-National Science Foundation,
and by a center of excellence supported by the Israel Science Foundation
(grant number 1665/10).
SBG is supported by the Golda Meir Foundation Fund.

\appendix
\section{Integrals}\label{app:integrals}

We shall consider here quivers coupled to a single pair of
messengers; see eq.~(\ref{wmgm}).
This can be generalized straightforwardly to a general messenger
sector, as discussed in section 2.
The sparticle masses are given by eq.~\eqref{eq:sparticle_masses},
where we presented the suppression functions
\beq
\mathcal{E}_i^t(x,\{y_\ell\}) \equiv
-\frac{8}{M^2 x^2 \, n(\mathbf{s}_{t}) } \int d^4 p \;
\frac{f_i^t(p^2)}{p^2}
\left[3\tilde{C}_1(p^2) -4\tilde{C}_{1/2}(p^2)
+\tilde{C}_0(p^2)
\right] \ ,
\label{eq:massfunction:app}
\eeq
where $M$ and $x$ are defined in eqs.~\eqref{wmgm} and
\eqref{eq:variables}, respectively.
The current correlators in eq.~(\ref{correnti}), in this case, are
given by \cite{Meade:2008wd}
\begin{align}
\tilde{C}_0^t(p^2) &=  \, n(\mathbf{s}_{t})  \int \frac{d^4 q}{(2\pi)^4}
\frac{1}{\left[q^2+m_+^2\right]\left[(p+q)^2+m_-^2\right]}
\ , \label{eq:C0} \\
\tilde{C}_{1/2}^t(p^2) &= -\frac{ \, n(\mathbf{s}_{t}) }{p^2} \int \frac{d^4 q}{(2\pi)^4}
\sum_{\pm}\left(\frac{1}{(p+q)^2+m_{\pm}^2}\right)
\frac{p\cdot q}{q^2+m_0^2} \ , \label{eq:C1/2} \\
\tilde{C}_{1}^t(p^2) &= -\frac{ \, n(\mathbf{s}_{t}) }{3p^2} \int \frac{d^4 q}{(2\pi)^4}
\bigg[\sum_{\pm}\left(\frac{(p+q)\cdot(p+2q)}
{\left[q^2+m_{\pm}^2\right]\left[(p+q)^2+m_{\pm}^2\right]}
-\frac{4}{q^2+m_{\pm}^2}\right) \non
&\phantom{= -\frac{2}{3p^2} \int \frac{d^4 q}{(2\pi)^4}\bigg[\ }
+\frac{4(p+q)\cdot q + 8m_0^2}
{\left[q^2+m_0^2\right]\left[(p+q)^2+m_0^2\right]} \bigg]
\ , \label{eq:C1}
\end{align}
where $m_0=M$ is the mass of the fermionic messengers and
$m_{\pm}^2=M^2 \pm F$ are the bosonic masses.
We define the following symbol \cite{vanderBij:1983bw}
\begin{align}
&\langle m_{11},\cdots,m_{1n_1}|m_{21},\cdots,m_{2n_2}|
m_{31},\cdots,m_{3n_3}\rangle \non
& \qquad
\equiv
\int \frac{d^d p \,d^d q}{\pi^d}
\prod_{i=1}^{n_1}\prod_{j=1}^{n_2}\prod_{k=1}^{n_3}
\frac{1}{p^2+m_{1i}^2}\frac{1}{q^2+m_{2j}^2}
\frac{1}{(p-q)^2+m_{3k}^2} \ .
\end{align}
Some useful results, evaluated in dimensional regularization with
$d=4-2 \epsilon$, are \cite{vanderBij:1983bw, Ghinculov:1994sd}
\begin{align}
\langle m_a|m_b|m_c\rangle &=
\frac{1}{-1+2\varepsilon}
\Big(m_a^2\langle m_a,m_a|m_b|m_c\rangle
+m_b^2\langle m_b,m_b|m_c|m_a\rangle \non
& \phantom{=\frac{1}{-1+2\varepsilon}\Big(\ }
+m_c^2\langle m_c,m_c|m_a|m_b\rangle\Big) \ , \\
\langle m_a,m_a|m_b|m_c\rangle &=
\frac{1}{2\varepsilon^2}
+\frac{1/2-\gamma-\log m_a^2}{\varepsilon}
+\gamma^2-\gamma+\frac{\pi^2}{12}
 \non & \phantom{=\ }
+\left(2\gamma-1\right)\log m_a^2 + \log^2 m_a^2 -\frac{1}{2}
+h(a,b) \ , \label{quattrone}
\end{align}
where we have defined the function
\begin{align}
h(a,b) &\equiv \int_0^1 \, dt
\left(1 + \Li_2(1-\mu^2) - \frac{\mu^2}{1-\mu^2}\log\mu^2\right) \ ,
\qquad
\mu^2 \equiv \frac{a t + b(1 - t)}{t(1 - t)} \ ,
\end{align}
with $a\equiv m_b^2/m_a^2$ and $b\equiv m_c^2/m_a^2$.
It turns out that the  integral in eq.~(\ref{quattrone}) is  infrared
divergent in the limit $m_a \rightarrow 0$;
in this limit we need then to introduce an infrared regulator mass
$m_\varepsilon$ \cite{Martin:1996zb} (which will drop out of the final result)
\begin{align}
\langle m_a|m_b|m_\varepsilon,m_\varepsilon\rangle &=
\frac{\Gamma(1+2\varepsilon)}{2}
\bigg(\frac{1}{\varepsilon^2}
+\frac{1-2\log m_\varepsilon^2}{\varepsilon} + 1
-\frac{\pi^2}{6}
-F_2(m_a^2,m_b^2)
-2F_3(m_a^2,m_b^2)
\non & \phantom{=\frac{\Gamma(1+2\varepsilon)}{2}
\bigg(\ }
+\left(-2+2F_1(m_a^2,m_b^2)\right)\log m_\varepsilon^2
+\log^2 m_\varepsilon^2\bigg) \ .
\end{align}
The functions above for $a\neq b$ are defined as \cite{Martin:1996zb}
\begin{align}
F_1(a,b) &\equiv \frac{a\log a - b\log b}{a - b} \ , \qquad
F_2(a,b) \equiv \frac{a\log^2 a - b\log^2 b}{a - b} \ , \non
F_3(a,b) &\equiv \frac{a\Li_2\left(1-\frac{b}{a}\right)
  -b\Li_2\left(1-\frac{a}{b}\right)}{a-b} \ ,
\end{align}
while
\begin{align}
F_1(a,a) \equiv 1 + \log a \ , \qquad
F_2(a,a) \equiv 2\log a + \log^2 a \ , \qquad
F_3(a,a) \equiv 2 \ .
\end{align}
Note that $h(0,b) = 1 + \Li_2(1-b)$.

Now we will introduce a formalism such that the integrals can
be carried out straightforwardly for any form factor $f(p^2)$.
Let us rewrite eq.~\eqref{eq:massfunction:app} as follows (suppressing
here for notational simplicity the indices $i,t$ for the nodes),
\begin{align}
&\mathcal{E}(x,\{y_\ell\}) =
\frac{1}{4 M^2 x^2}
\int \frac{d^4p\, d^4q}{\pi^4} f(p^2)
\bigg[
-\frac{2}{p^2\left[q^2+m_+^2\right]\left[(p+q)^2+m_-^2\right]}
\non &
+\sum_{\pm}\frac{4}{p^2\left[q^2+m_0^2\right]\left[(p+q)^2+m_{\pm}^2\right]}
-\sum_{\pm}\frac{1}{p^2\left[q^2+m_{\pm}^2\right]\left[(p+q)^2+m_{\pm}^2\right]}
\non &
-\frac{4}{p^2\left[q^2+m_0^2\right]\left[(p+q)^2+m_0^2\right]}
+\sum_{\pm}\frac{4\left(m_{\pm}^2 -
  m_0^2\right)}{p^4\left[q^2+m_0^2\right]\left[(p+q)^2+m_{\pm}^2\right]}
\non &
-\sum_{\pm}\frac{4m_{\pm}^2}{p^4\left[q^2+m_{\pm}^2\right]\left[(p+q)^2+m_{\pm}^2\right]}
+\frac{8m_0^2}{p^4\left[q^2+m_0^2\right]\left[(p+q)^2+m_0^2\right]}
\bigg] \ ,
\end{align}
where the sum over $\pm$ is understood as the sum over the masses
$m_+$ and $m_-$.
It is easily seen that the terms can be split into two classes.
The first comprises the first four terms by having only one massless
propagator $1/p^2$, and we will denote this class the $\alpha$ terms,
while the second class has two massless propagators $1/p^4$, and
correspondingly we will denote this class the $\beta$ terms.
Using the fact that all the form factor functions $f(p^2)$ can be
expanded in partial fractions for the $\alpha$ terms as
\beq
\frac{f(p^2)}{p^2} = \frac{a_0}{p^2}
+\sum_\ell \frac{a_{1,\ell}}{p^2 + m_\ell^2}
+\sum_\ell \frac{a_{2,\ell}}{\left(p^2 + m_\ell^2\right)^2} \ ,
\label{eq:pf_alpha}
\eeq
where the sum is over all the mass poles not counting multiplicity,
while for the $\beta$ terms the corresponding partial fractions read
\beq
\frac{f(p^2)}{p^4} = \frac{b_{-1}}{p^2} + \frac{b_0}{p^4}
+\sum_\ell \frac{b_{1,\ell}}{p^2 + m_\ell^2}
+\sum_\ell \frac{b_{2,\ell}}{\left(p^2 + m_\ell^2\right)^2} \ ,
\label{eq:pf_beta}
\eeq
we can proceed as follows. First, we define the basic functions
\begin{align}
\alpha_0(x) &\equiv \frac{1}{4M^2}\Big(
\sum_{\pm}\left[ 4\langle 0 | m_0 | m_\pm\rangle
  -\langle 0 | m_\pm | m_\pm\rangle \right]
-2\langle 0 | m_+ | m_-\rangle
-4\langle 0 | m_0 | m_0\rangle \Big) \ , \non
\alpha_1(x,y_\ell) &\equiv \frac{1}{4M^2}\Big(
\sum_{\pm}\left[ 4 \langle m_\ell | m_0 | m_\pm\rangle
  -\langle m_\ell | m_\pm | m_\pm\rangle \right]
-2\langle m_\ell | m_+ | m_-\rangle
-4\langle m_\ell | m_0 | m_0\rangle \Big) \ , \non
\alpha_2(x,y_\ell) &\equiv \frac{1}{4}\Big(
\sum_{\pm}\left[ 4 \langle m_\ell , m_\ell | m_0 | m_\pm\rangle
  -\langle m_\ell , m_\ell | m_\pm | m_\pm\rangle \right]
-2\langle m_\ell , m_\ell| m_+ | m_-\rangle
\non & \phantom{\equiv \frac{1}{4M^4}\Big(\ }
-4\langle m_\ell , m_\ell | m_0 | m_0\rangle \Big) \ , \label{aaabbb}\\
\beta_{-1}(x) &\equiv \frac{1}{M^4} \Big(
\sum_{\pm}\left[\left(m_\pm^2-m_0^2\right)\langle 0 | m_0 | m_\pm
\rangle
-m_\pm^2\langle 0 | m_\pm | m_\pm \rangle \right]
+2m_0^2\langle 0 | m_0 | m_0 \rangle \Big) - r(x) \ , \non
\beta_0(x) &\equiv \frac{1}{M^2}\Big(
\sum_{\pm}\left[\left(m_\pm^2-m_0^2\right)\langle 0 , 0 | m_0 | m_\pm
\rangle
-m_\pm^2\langle 0 , 0 | m_\pm | m_\pm \rangle \right]
+2m_0^2\langle 0 , 0 | m_0 | m_0 \rangle
\Big)\ , \non
\beta_1(x,y_\ell) &\equiv \frac{1}{M^4} \Big(
\sum_{\pm}\left[\left(m_\pm^2-m_0^2\right)\langle m_\ell | m_0 | m_\pm
\rangle
-m_\pm^2\langle m_\ell | m_\pm | m_\pm \rangle \right]
+2m_0^2\langle m_\ell | m_0 | m_0 \rangle
\Big) - r(x) \ ,
\nonumber
\end{align}
\begin{align}
\beta_2(x,y_\ell) &\equiv \frac{1}{M^2}\Big(
\sum_{\pm}\left[\left(m_\pm^2-m_0^2\right)\langle m_\ell , m_\ell | m_0 | m_\pm
\rangle
-m_\pm^2\langle m_\ell , m_\ell | m_\pm | m_\pm \rangle \right]
\non & \phantom{\equiv \frac{1}{M^2}\Big(\ }
+2m_0^2\langle m_\ell , m_\ell | m_0 | m_0 \rangle
\Big)\ . \nonumber
\end{align}
We have not presented the function $r(x)$ in $\beta_{-1}$ and $\beta_1$ in eq. (\ref{aaabbb}),
since it must drop out of the final result
because of infrared cancellations. Hence, one should check that (see eq. (\ref{eq:Efunc_def}) below)
\beq
b_{-1} + \sum_\ell b_{1,\ell} = 0 \ ,
\eeq
and since $r(x)$ is not a function of $y_\ell$ it will indeed drop out.

Carrying out the integrals yields
\begin{align}
\alpha_0(x) &=
-\Li_2(-x) - (1+x)\Li_2\left(\frac{x}{1+x}\right)
+\frac{1}{2}(1+x)\Li_2\left(\frac{2x}{1+x}\right)
+(x \to -x) \ , \label{eq:alpha0} \\
\alpha_1(x,y) &=
\h{y^2}{1}
-\h{y^2}{1+x}
+\frac{y^2}{2}\h{\frac{1}{y^2}}{\frac{1}{y^2}}
-y^2\h{\frac{1+x}{y^2}}{\frac{1}{y^2}}
\non & \phantom{=\ }
+\frac{y^2}{4}\h{\frac{1+x}{y^2}}{\frac{1-x}{y^2}}
+\frac{y^2}{4}\h{\frac{1+x}{y^2}}{\frac{1+x}{y^2}}
+\frac{1+x}{2}\h{\frac{y^2}{1+x}}{1}
\non & \phantom{=\ }
-(1+x)\h{\frac{y^2}{1+x}}{\frac{1}{1+x}}
+\frac{1+x}{2}\h{\frac{y^2}{1+x}}{\frac{1-x}{1+x}}
+(x \to -x) \ , \label{eq:alpha1}\\
\alpha_2(x,y) &=
-\frac{1}{2}\h{\frac{1}{y^2}}{\frac{1}{y^2}}
+\h{\frac{1+x}{y^2}}{\frac{1}{y^2}}
-\frac{1}{4}\h{\frac{1+x}{y^2}}{\frac{1+x}{y^2}}
-\frac{1}{4}\h{\frac{1+x}{y^2}}{\frac{1-x}{y^2}}
\non & \phantom{=\ }
+(x \to -x) \ , \label{eq:alpha2}
\end{align}
\begin{align}
\beta_{-1}(x) &=
\frac{x^2}{2}
-x\Li_2(-x) - (1+x)x\Li_2\left(\frac{x}{1+x}\right)
+(x \to -x) \ , \label{eq:betam1}\\
\beta_0(x) &=
(1+x)\log(1+x) + \Li_2(-x)
-(1+x)\Li_2\left(\frac{x}{1+x}\right)
+(x \to -x) \ , \label{eq:beta0}\\
\beta_1(x,y) &=
-2\h{y^2}{1}
-x\h{y^2}{1+x}
-y^2\h{\frac{1}{y^2}}{\frac{1}{y^2}}
-x y^2\h{\frac{1+x}{y^2}}{\frac{1}{y^2}}
\non & \phantom{=\ }
+(1+x)y^2\h{\frac{1+x}{y^2}}{\frac{1+x}{y^2}}
+2(1+x)^2\h{\frac{y^2}{1+x}}{1}
\non & \phantom{=\ }
-(1+x)x\h{\frac{y^2}{1+x}}{\frac{1}{1+x}}
-\frac{x^2}{2}
+(x \to -x) \ , \label{eq:beta1} \\
\beta_2(x,y) &=
\h{\frac{1}{y^2}}{\frac{1}{y^2}}
+x\h{\frac{1+x}{y^2}}{\frac{1}{y^2}}
-(1+x)\h{\frac{1+x}{y^2}}{\frac{1+x}{y^2}}
+(x \to -x) \ . \label{eq:beta2}
\end{align}
Finally, using these definitions along with the coefficients of the partial
fractions defined in eqs.~\eqref{eq:pf_alpha} and \eqref{eq:pf_beta},
we can conveniently write
\begin{align}
\label{eq:Efunc_def}
\mathcal{E}(x,\{y_\ell\}) &= \frac{1}{x^2}
\Big(
a_0 \alpha_0(x)
+ b_{-1} M^2 \beta_{-1}(x)
+ b_0 \beta_0(x)
\\ & \phantom{= \frac{1}{x^2}\ }
+ \sum_\ell\left[
a_{1,\ell} \alpha_1(x,y_\ell)
+ \frac{a_{2,\ell}}{M^2}\,  \alpha_2(x,y_\ell)
+ b_{1,\ell} M^2 \beta_1(x,y_\ell)
+ b_{2,\ell} \beta_2(x,y_\ell)
\right]
\Big) \ , \nonumber
\end{align}
where as already mentioned, the sum is over all the mass poles in
$f(p^2)$ not counting multiplicity.

\subsection{Example: MGM}

As a simple example, we can calculate the suppression function for minimal
gauge mediation as follows: $f(p^2) = 1$ implies that the only
non-zero coefficients of the partial fractions are $a_0 = b_0 = 1$ and
hence the result is \cite{Dimopoulos:1996gy,Martin:1996zb}
\beq
\mathcal{E}(x) = \frac{1}{x^2}
\left(\alpha_0(x) + \beta_0(x)\right) \ .
\label{eq:Efunc_MGM}
\eeq
This function approaches $1$ for small $x$,
as will be discussed in the next sub-appendix.

\subsection{Limits}\label{app:limits}

Let us first consider the limit $x\to 0$, which corresponds to the
messenger scale being much larger than the SUSY-breaking scale:
$M\gg \sqrt{F}$. It is also motivated by the fact that the soft masses
do not vary much for $x\lesssim 0.7$, and hence it is a good
approximation for a large range. Since all the functions are even in
$x$ and they are all multiplied by $1/x^2$, the only term we need to
calculate is that of order $x^2$ in the Taylor expansion.
For the $\alpha$ class functions we get
\beq
\lim_{x\to 0} \frac{\alpha_0(x)}{x^2} =
\lim_{x\to 0} \frac{\alpha_1(x,y)}{x^2} =
\lim_{x\to 0} \frac{\alpha_2(x,y)}{x^2} = 0 \ ,
\eeq
while for the $\beta$ class functions we obtain
\begin{align}
\lim_{x\to 0} \frac{\beta_{-1}(x)}{x^2} &=
\lim_{x\to 0} \frac{\beta_0(x)}{x^2} = 1 \ , \\
\tilde{\beta}_1(y) \equiv \lim_{x\to 0} \frac{\beta_1(x,y)}{x^2} &=
-1 + 2\h{y^2}{1}
-2\iotafunc{y^2}{1}{y^2}{0}
+2y^2\iotafunc{\frac{1}{y^2}}{\frac{1}{y^2}}{\frac{1}{y^2}}{0}
\non &\phantom{=\ }
+y^2\sigmafunc{\frac{1}{y^2}}{\frac{1}{y^2}}{\frac{1}{y^2}}{\frac{1}{y^2}}
+2\sigmafunc{y^2}{1}{y^2}{0} \ , \label{eq:x0limitbeta1}\\
\tilde{\beta}_2(y) \equiv \lim_{x\to 0} \frac{\beta_2(x,y)}{x^2} &=
-2\iotafunc{\frac{1}{y^2}}{\frac{1}{y^2}}{\frac{1}{y^2}}{0}
-\sigmafunc{\frac{1}{y^2}}{\frac{1}{y^2}}{\frac{1}{y^2}}{\frac{1}{y^2}}
\ , \label{eq:x0limitbeta2}
\end{align}
where we have defined the following functions
\begin{align}
\iotafunc{a}{b}{c}{d} &\equiv
- \int_0^1 \frac{dt}{1-\mu^2}
\left(1+\frac{\mu^2}{1-\mu^2}\log \mu^2\right) \nu \ , \qquad
\mu^2 \equiv \frac{a}{1-t} + \frac{b}{t} \ , \qquad
\nu \equiv \frac{c}{1-t} + \frac{d}{t} \ , \non
\sigmafunc{a}{b}{c}{d} &\equiv
- \int_0^1 \frac{dt}{\left(1-\mu^2\right)^2}
\left(2+\frac{1+\mu^2}{1-\mu^2}\log \mu^2\right) \nu^2 \ ,
\end{align}
and the expansion of the function $h(a(x),b(x))$ is made as follows
\begin{align}
h(a(x),b(x)) &= h(a(0),b(0))
+ \iotafunc{a(0)}{b(0)}{a'(0)}{b'(0)} x
\\ &\phantom{=\ }
+ \frac{1}{2}\sigmafunc{a(0)}{b(0)}{a'(0)}{b'(0)} x^2
+ \frac{1}{2}\iotafunc{a(0)}{b(0)}{a''(0)}{b''(0)} x^2
+ \mathcal{O}(x^3) \ , \nonumber
\end{align}
where $a'(x) = d a(x)/dx$, etc.

There is another limit, which merely serves as a check of the
calculations, i.e.~taking all $y$'s to infinity.
Using that
\begin{align}
\lim_{y\to\infty} \h{a y^2}{X} &= 1 + \Li_2(-a y^2) \ , \\
\h{\frac{a}{y^2}}{\frac{b}{y^2}} &= 1 + \frac{\pi^2}{6}
- \frac{a+b}{y^2}\left(1 + \log y^2\right)
+ \frac{a}{y^2} \log a
+ \frac{b}{y^2} \log b + \mathcal{O}(y^{-4}) \ ,
\end{align}
we find that
\begin{align}
\lim_{y\to\infty} \alpha_1(x,y) =
\lim_{y\to\infty} y^2\alpha_2(x,y) =
\lim_{y\to\infty} \frac{\beta_{-1}(x)}{y^2} =
\lim_{y\to\infty} \frac{\beta_1(x,y)}{y^2} =
\lim_{y\to\infty} \beta_2(x,y) = 0 \ ,
\end{align}
where the factors of $y^2$ multiplying the functions are always present
for dimensional reasons. Then it is easy to see that any mass factor
reduces to
\beq
\mathcal{E}(x,\infty) = \frac{1}{x^2}
\left(a_0 \alpha_0(x) + b_0 \beta_0(x)\right) \ .
\eeq
Hence, if $a_0 = b_0 = 1$ the result reduces to that of MGM.
This is indeed the case for all the masses calculated
in this note (independent of which node the field is charged under).

A final limit, which is a bit harder to calculate, is the limit of
$y\to 0$. This limit serves only as a consistency check and hence we
will not give all the details. The result after the dust has settled
is
\begin{align}
\lim_{y\to 0} \alpha_1(x,y) = \alpha_0(x) \ , \qquad
\lim_{y\to 0} y^2\alpha_2(x,y) = 0 \ , \qquad
\lim_{y\to 0} \beta_2(x,y) = \beta_0(x) \ ,
\end{align}
while expanding we find
\beq
\beta_1(x,y) = \beta_{-1}(x) - y^2 \beta_0(x) + \mathcal{O}(y^3) \ .
\eeq
Thus eq.~\eqref{eq:Efunc_def} simplifies in this limit as follows
\begin{align}
\mathcal{E}(x,0) = \frac{1}{x^2}\left[
\left(a_0 + \sum_\ell a_{1,\ell}\right) \alpha_0(x)
+\left(b_0 + \sum_\ell\left(b_{2,\ell} -
\tilde{b}_{1,\ell}\right)\right)\beta_0(x) \right] \ ,
\end{align}
where we have defined
$\tilde{b}_{1,\ell} \equiv \lim_{y_\ell\to 0} M^2 y_\ell^2 b_{1,\ell}$.
This result is after all expected and one can check that the formula
gives zero for all nodes which are not connected to the messengers
while it gives the MGM result \eqref{eq:Efunc_MGM}
multiplied by $g_i^4/g_{\rm eff}^4$ with $i$ being the node under
consideration.

\end{document}